\documentclass[acmsmall]{acmart}\usepackage[]{graphicx}\usepackage[]{color}
\makeatletter
\def\maxwidth{ %
  \ifdim\Gin@nat@width>\linewidth
    \linewidth
  \else
    \Gin@nat@width
  \fi
}
\makeatother

\definecolor{fgcolor}{rgb}{0.345, 0.345, 0.345}

\usepackage{framed}
\makeatletter
 {\par\unskip\endMakeFramed%
 \at@end@of@kframe}
\makeatother

\definecolor{shadecolor}{rgb}{.97, .97, .97}
\definecolor{messagecolor}{rgb}{0, 0, 0}
\definecolor{warningcolor}{rgb}{1, 0, 1}
\definecolor{errorcolor}{rgb}{1, 0, 0}

\usepackage{alltt}

\usepackage[utf8]{inputenc}

\usepackage[T1]{fontenc}
\usepackage{textcomp}

\usepackage{graphicx}
\usepackage{enumerate}

\usepackage[american]{babel}

\def\citepos#1{\citeauthor{#1}'s (\citeyear{#1})}

\AtBeginDocument{%
  \providecommand\BibTeX{{%
    \normalfont B\kern-0.5em{\scshape i\kern-0.25em b}\kern-0.8em\TeX}}}

\def\citepos#1{\citeauthor{#1}'s \cite{#1}}

\setcopyright{rightsretained}
\IfFileExists{upquote.sty}{\usepackage{upquote}}{}
\begin{document}

\title{Taboo and Collaborative Knowledge Production: Evidence from Wikipedia}

\author{Kaylea Champion}
\affiliation{%
  \institution{University of Washington}
  \city{Seattle, WA}
  \country{United States}
}
\email{kaylea@uw.edu}

\author{Benjamin Mako Hill}
\affiliation{%
  \institution{University of Washington}
  \city{Seattle, WA}
  \country{United States}
}
\email{makohill@uw.edu}

\begin{abstract}

By definition, people are reticent or even unwilling to talk about taboo subjects. Because subjects like sexuality, health, and violence are taboo in most cultures, important information on each of these subjects can be difficult to obtain. Are peer produced knowledge bases like Wikipedia a promising approach for providing people with information on taboo subjects? With its reliance on volunteers who might also be averse to taboo, can the peer production model produce high-quality information on taboo subjects? In this paper, we seek to understand the role of taboo in knowledge bases produced by volunteers. We do so by developing a novel computational approach to identify taboo subjects and by using this method to identify a set of articles on taboo subjects in English Wikipedia. We find that articles on taboo subjects are more popular than non-taboo articles and that they are frequently vandalized. Despite frequent vandalism attacks, we also find that taboo articles are higher quality than non-taboo articles. We hypothesize that stigmatizing societal attitudes will lead contributors to taboo subjects to seek to be less identifiable. Although our results are consistent with this proposal in several ways, we surprisingly find that contributors make themselves more identifiable in others.

\end{abstract}

\acmJournal{PACMHCI}
\acmYear{2023} \acmVolume{7} \acmNumber{CSCW2} \acmArticle{299} \acmMonth{10} \acmPrice{}\acmDOI{10.1145/3610090}
\makeatletter
\gdef\@copyrightpermission{
  \begin{minipage}{0.2\columnwidth}
   \href{https://creativecommons.org/licenses/by/4.0/}{\includegraphics[width=0.90\textwidth]{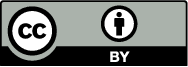}}
  \end{minipage}\hfill
  \begin{minipage}{0.8\columnwidth}
   \href{https://creativecommons.org/licenses/by/4.0/}{This work is licensed under a Creative Commons Attribution International 4.0 License.}
  \end{minipage}
  \vspace{5pt}
}
\makeatother

\maketitle

\section{Introduction}
Taboos are behavioral prohibitions characterized by the notion that violations of the prohibition create symbolic uncleanness or pollution \citep{allan_taboo_2019, douglas_purity_1978}. Although taboos vary enormously across cultures, they exist in virtually all societies \citep{allan_taboo_2019, douglas_purity_1978}. People try to avoid being polluted by taboo in a range of ways---frequently by not talking about or even mentioning taboo subjects. When taboo subjects are discussed, people refer to them indirectly or vaguely. 
As a result, access to high-quality information on taboo subjects is frequently difficult.
This is problematic because taboo subjects often include important subjects such as mental health, reproduction, menstruation, abuse, extremism, and human rights violations.

The growth of the Internet and the development of free high-quality knowledge bases like Wikipedia have made a wealth of information widely available. Are Internet knowledge bases a promising approach to providing people with information on taboo subjects? In that they provide opportunities for privacy in information consumption, social computing systems have been cited as promising avenues for accessing information on a range of taboo subjects like menstruation \citep{sondergaard_designing_2021,  campo_woytuk_touching_2020, tuli_learning_2018, almeida_looking_2016}, sexual abuse, and harassment \citep{moitra_understanding_2020, andalibi_understanding_2016}. On the other hand, many online knowledge bases rely on volunteers who choose their own tasks and subject areas, and it seems reasonable to assume that at least some volunteers may be reticent to contribute to resources about taboo subjects.
In either case, there is reason to believe that taboo shapes knowledge production in social computing systems. That said, we know of no work that explores the role that taboo plays in online knowledge production.

One challenge to studying taboo as a more general phenomenon is to develop a way to identify taboos systematically. Examining taboo from a systematic perspective allows us to develop evidence about taboo as a broad social phenomenon. To do so, we draw inspiration from work in linguistics to develop a novel computational approach \citep{burridge_taboo_2015}. Our method uses a supervised machine learning classifier trained on words in dictionary definitions associated with euphemisms---i.e., subjects that speakers go out of their way to describe indirectly, often at the cost of clarity. 
To understand how taboo might be shaping activity in social computing, we use these words from dictionary definitions to identify a set of articles on taboo subjects in English Wikipedia as well as a comparison set of otherwise similar articles. 
For example, since the term ``passed away'' is a euphemism that would include the word ``dead'' or ``death'' in its definition, we might identify the Wikipedia article on ``death'' as taboo.

Using detailed digital trace data on articles and contributors from Wikipedia, we test five hypotheses derived from theory about how taboo will change the way that Wikipedia articles are consumed and produced. 
We find support for hypotheses that articles on taboo subjects are more popular than non-taboo articles and that taboo articles are more frequently vandalized than non-taboo articles. Despite these frequent attacks, we also find that taboo articles are of higher quality than non-taboo articles, contrary to our hypothesis. We hypothesize that societal attitudes against contact or association with these subjects would lead contributors to taboo subjects to seek to be less identifiable, and although this is sustained in part by our results, it is also contradicted in part.
These surprising results suggest that contributors to the public production of taboo knowledge navigate more complex privacy trade-offs than previously theorized.

The remaining sections of the paper are structured as follows. In §\ref{sec:background}, we examine theories of peer production, taboo, social norms, and identifiability/anonymity, including what these theories suggest about the peer production of information about taboo subjects. In §\ref{sec:setting}, we describe our empirical setting Wikipedia. We then describe the novel technique we develop to identify taboo and the construction of our analytic sample in §\ref{sec:methods}, and our analytical process in §\ref{sec:plan}. Section §\ref{sec:results} reports results from our hypothesis tests. We discuss the significance and implications of our findings in §\ref{sec:discussion}, some important limitations of this work in §\ref{sec:limitations},
and conclude in §\ref{sec:conclusion}.

\section{Background}
\label{sec:background}
\subsection{Commons-Based Peer Production}
As \citet{benkler_coases_2002} describes, one of the technological and social advances of the last few decades is the creation of a novel form of production: commons-based peer production. Individuals participating in commons-based peer production activities online are volunteers who self-organize around a goal, select their own tasks, and engage in the creation or maintenance of information goods. 
These volunteers are sometimes motivated by generosity and altruism, or simply pursuit of their own passions---which may be in contradiction to dominant social and market logics. 
Often these information goods are also public goods, as in the case of Wikimedia projects such as Wikipedia and Wiktionary, the mapping project Open Street Map, and free/libre open source software such as GNU/Linux and the Python programming language. These valuable projects are critical parts of our digital infrastructure: delivering knowledge that answers our searches online, the software that organizes and displays those answers, and the servers and protocols that convey them from distant data centers to our fingertips. Although these projects are rightfully hailed as transformative, their results---and the peer production process that produces them---are far from perfect.

Studies of peer production projects have pointed out numerous examples of bias and neglect. The participant pool of many prominent examples of peer production is generally male, including Wikipedia \cite{ford_anyone_2017,hill_wikipedia_2013,collier_conflict_2012}, Linux \cite{reagle_free_2013,nafus_patches_2012}, and Open Street Map \cite{thebault-spieker_geographic_2018}. The products developed through peer production may also be biased or neglect important materials. In Wikipedia, previous work has found neglect of articles about women \cite{tripodi_ms_2021}, non-English languages \cite{khatri_social_2022,miquel-ribe_wikipedia_2020,hecht_tower_2010}, and countries, religion, and LGBTQ subjects \cite{warncke-wang_misalignment_2015}.
In Open Street Map, research has found a lack of map information about the global South \cite{thebault-spieker_geographic_2018}. 
These biases in peer-produced resources are troubling because they reflect and may serve to perpetuate existing societal biases, inequalities, and hegemonic structures. 
Despite these flaws, however, peer production has such advantages as self-organization and self-selection of tasks.

Examining the question of taboo knowledge in peer production offers us insight into the extent to which this novel organizational form is simply serving to reproduce and magnify existing features of society, or whether participants may perhaps be finding ways to resist cultural norms.

\subsection{Taboo---and Taboo in HCI} 

Taboo has been the subject of an enormous amount of scholarship in anthropology, sociology, and linguistics \citep{allan_taboo_2019, douglas_purity_1978}.
Taboos demarcate the forbidden and unspeakable parts of existence from those that are recognized as sanctified, worthy, or simply acceptable. A taboo acts as a behavioral prohibition and is characterized by the notion that violations of the taboo create a sense of uncleanness or pollution. The uncleanness can be literal (e.g., contact with something covered in germs may transfer those germs) or may be symbolic (e.g., make one's prayers unacceptable in the eyes of God). 
Because making contact with something taboo makes a person unclean, taboo spreads its uncleanness through interaction. 
Although the taboo itself may make someone feel personally uncomfortable or embarrassed even if they are alone, taboo is often enacted socially. Although speakers may use euphemism to protect the sensibilities of others when discussing taboo subjects, \citet{mcglone_looking_2003} found through an experimental manipulation that speakers are more likely to use euphemism when discussing taboo subjects if they expect to be identified to the listener, concluding that the reticence to speak about taboo subjects is more a matter of a speaker trying to save face than protecting others. Taboo may manifest in complex ways in social computing settings. We may not only have a sense of being visible to friends, but also imagine a broader audience, which might be profoundly public \citep{litt_knock_2012,marwick_i_2011}. We are further visible to our technology---not only in the form of history and device logs, but also in the ways our behaviors train the algorithms that curate our experience---any of which may serve to embarrass us, restrain us, and drive us to seek privacy \cite{toch_personalization_2012}.

Mary Douglas's work in \textit{Purity and Danger: An Analysis of the Concepts of Pollution and Taboo} points out that when ``pollution rules'' that constitute taboo are examined with care, social order is constantly implicated in what we consider clean or unclean, acceptable or forbidden. Ultimately, she says, we should understand taboo as part of a symbolic system in which ``uncleanness is matter out of place'' \citep[p.~41]{douglas_purity_1978} and a violation of ``cherished classifications'' (\textit{ibid}, p. 37). Hence, to the extent that people interact with subjects that are taboo, they are in some way challenging social order and by extension society itself. As a result, reference materials associated with taboo subjects may be restricted, censored, or banned outright \citep{allan_taboo_2019}. 

The English word ``taboo'' originated in the voyages of Captain Cook who derived the word from the Tongan word ``tapu'' meaning sacred or forbidden. The term came to be used in colonial and anthropological accounts of non-Western cultures, often as part of characterizing those cultures as ``uncivilized'' \citep{allan_taboo_2019}. 
Despite this history, taboo has come to be understood as existing in virtually all cultures and societies throughout history \cite{allan_taboo_2019,crespo_fernandez_conceptual_2011,douglas_purity_1978}. 
Although every member of every society has some internal and personal sense of taboo, taboos vary by culture, religion, relative levels of privilege, and more. Like early colonial anthropologists, we may fail to recognize our own taboos because they seem natural or even objective to us \citep{douglas_purity_1978}. Seeking out a more systematic way to identify taboo may therefore serve to de-center and de-naturalize our own cultural position.

Despite the rich literature in other disciplines, we believe that ours is the first article in social computing to approach the study of taboo in general, instead of focusing on specific taboo subjects.
To what extent is ``taboo'' a subject of concern for research at the intersection of society, technology, and design? 
Although a search of the over 2.9 million articles in the ACM digital library returns 1,849 results for `taboo', most of these articles refer to an algorithmic search strategy in which potential solutions are marked as forbidden or to a party game in which players try to guess a word when prompted with clues that cannot include the word itself.
A closer examination of the 174 research articles published in SIGCHI venues revealed that most of the remaining articles mentioned the term ``taboo'' only once and in passing.

Although none of these articles sought to understand or measure the effect of taboo in general, our search did reveal a number of articles on systems and designs that are intended to
counter or transcend the harmful effects of taboo on information environments in specific cases such as menstruation \citep{sondergaard_designing_2021,  campo_woytuk_touching_2020, almeida_looking_2016}, sexuality \citep{kannabiran_how_2011, rahman_adolescentbot_2021}, 
toilet training \citep{helms_you_2019}, and AIDS education \citep{sorcar_sidestepping_2017}.
Another line of research seeks to understand how people use existing social computing systems in ways that are shaped by specific taboos---e.g., sharing information about menstrual health \citep{tuli_learning_2018} and menopause \citep{lazar_parting_2019},  
or describing experiences such as sexual abuse \citep{andalibi_understanding_2016}, sexual harassment \citep{moitra_understanding_2020} and pregnancy loss \citep{andalibi_disclosure_2020}. 

One related body of work is the study of stigma, which is a connected but distinct concept. A taboo is a behavioral prohibition, whereas stigma refers to an identity state, perhaps but not always associated with violating taboos. One may be stigmatized for a congenital condition or by violating a range of different types of social norms. In \textit{Stigma: Notes on the Management of Spoiled Identity}, Erwin Goffman describes stigma as ``an undesired differentness from what is expected'' \citep[p.~5]{goffman_stigma_1963} and, like Douglas, invokes the role of social order. For Goffman, a stigmatized identity means that one's social identity as perceived externally is in some way different from what is perceived internally. For example, a formerly incarcerated person may be stigmatized---they see themselves as `like everyone else' and trying to live a normal life, but to their neighbors they may forever be discredited as criminal. Goffman observes that stigma in society is transferred via association: the child of a person known to be a criminal may also be the subject of suspicion.

Although none of this work takes up the concept of taboo in general, these papers point to the fact that taboo is an important feature of online information seeking and production in at least several specific cases. %
Taking a systematic approach is complementary to these studies that focus on \textit{a priori} taboos by supporting comparative analysis across contexts and across a range of taboos, as well as potentially expanding the set of taboos explored through various methods.

\subsection{Taboo Content Consumption}

Avoidance of taboo is rarely universal. We may simply need information related to taboo subjects, such as those related to such everyday activities as urination, defecation, menstruation, and sexual intercourse. 
Some people deliberately violate taboos and seek out community in doing so. Others are struggling with taboo issues or have survived taboo experiences and are seeking supportive environments. Further, the forbidden nature of taboos may lead some people to feel drawn toward the subject---they may titillate or fascinate or offer an avenue to rebel against society's norms. 

Privacy may also play a role in taboo violation. For example, \citet{malinowski_crime_1926} famously observed in his 1926 account of the people of Trobriand Island that a taboo may be widely violated in private and the subject of gossip and humor, yet still have severe consequences when its violation is turned into a public spectacle. 
Likewise, support groups often have rules about maintaining participant anonymity, and some sites have norms and affordances that support anonymized interactions about taboo subjects \citep{ammari_self-declared_2019}.

Web browsing may feel relatively private, especially when we are only consuming content. We may feel less inhibited in an online environment and more willing to engage with material that we would be reluctant to be seen reading about or heard speaking about publicly \citep{suler_online_2004}. One might expect that feelings of safety to be higher in websites---like Wikipedia---that are accessible without paywalls or registration requirements which signal to a reader that consumption is being tracked. 
Given that information seekers may have relatively few alternative sources for high-quality information about taboo subjects and the relative privacy of reading Wikipedia, we suggest \textbf{H1: peer produced resources about taboo subjects will receive higher readership than peer produced resources about other comparable subjects}.

\subsection{Taboo, Public Knowledge, and Identifiability}

Although consuming content hosted in knowledge bases like Wikipedia may feel private, producing knowledge resources is often a publicly visible act.
In Wikipedia, every article has a history tab containing each past version of itself. A ``diff'' view allows anyone to examine the letter-by-letter impact of every change made. Beyond the contribution itself, the account or IP address of every contributor is tracked and made visible. Both the article's revision history and an individual contributor's contribution history can be reviewed by anyone. Contributors can offer additional information about themselves by customizing their user page, by registering an email address that can be used to contact them, and by disclosing their gender.
Although the effort required to identify contributors varies, Wikipedia's high level of transparency means that every contributor is identifiable in some dimension.
Users of tools to protect personal privacy, including anonymity-preserving proxies such as the popular Tor browser, are banned from contributing to Wikipedia \citep{tran_are_2020}.

On Wikipedia, people edit only what they want to edit \cite{benkler_wealth_2006}. Because taboos are so powerful and pervasive, they may become internalized. As a result, contributors may seek to avoid taboo subjects even in private settings. 
A high level of public visibility and the voluntary nature of peer produced projects suggest that people may be reluctant to choose to contribute to taboo subjects out of a reluctance to contaminate themselves. Therefore, we propose
\textbf{H2: peer produced resources about taboo subjects will receive fewer contributions than other comparable peer produced subjects}.

Online knowledge bases are typically secondary or tertiary sources of information that depend on the availability of other sources to establish credibility \cite{ford_getting_2013}.
Taboo subjects are often the targets of censorship efforts \cite{allan_taboo_2019}. For example, materials about taboo subjects like violence, sexuality, race, and religion are regular targets of book banning campaigns in K-12 schools in the United States \cite{aucoin_censorship_2021}. Blocking of political, religious, and social norm-violating content (including pornography, drugs, LGBTQ content, and online dating) is widespread \cite{zittrain_shifting_2017}. At the state level, for example, China suppresses information on human rights abuses and collective action, 
and Norway blocks many pornography and gambling websites \cite{sundara_raman_censored_2020, zittrain_shifting_2017, king_how_2013}. 
Although social pressure and a lack of available secondary sources might lead to lower-quality contributions, the nature of taboo subjects themselves may tend to draw in or inspire bad faith contributions---e.g., writing sexual slurs on pages about sexual subjects.
Given societal and governmental opposition to materials on taboo subjects, the suppression of primary and secondary sources that might otherwise be used to build knowledge bases, and the role of troll, griefers, and vandals, we propose  
\textbf{H3: peer produced resources about taboo subjects will receive lower-quality contributions than peer produced resources about other comparable subjects}.

Some theorists of open content production have made direct connections between the quality of a resource and the volume of contributors.  \citet{raymond_cathedral_2001} famously described this process as ``Linus's Law:'' ``given enough eyeballs, all bugs are shallow.'' Empirically, \citet{haklay_how_2010} found that positional accuracy in open mapping projects increases with the number of participants. \citet{greenstein_open_2016} found that neutrality in point of view in Wikipedia articles improved with increases in the number of contributors. 
Synthesizing these observations---and as a consequence of H2 (fewer overall contributions) and H3 (lower-quality contributions)---we propose
\textbf{H4: peer produced resources about taboo subjects will be lower quality than peer produced resources about other comparable subjects}. 

Finally, we consider the impact of taboo on the willingness of contributors to identify themselves. Possessing knowledge about a taboo subject may lead to stigma \citep{goffman_stigma_1963}. 
Social embarrassment and opprobrium can result from being known to have violated taboos leading to loss of reputation and position, or in extreme cases, to incarceration and violence. 
For example, Wikipedia editors have been harassed for their contributions both on and off Wikipedia \cite{jacobs_wikipedia_2019, brown_female_2016, paling_wikipedias_2015}. In some cases, editors have even been prosecuted for their edits, as in the case of the Belarusian government arresting a Wikipedian for writing that Russia has invaded Ukraine \cite{song_top_2022}. 
Hence, people contributing knowledge about a taboo subject to online knowledge bases may engage in a range of strategies to control information about their identity and behavior \citep{semaan_impression_2017}.
For example, the Russian language edition of Wikipedia has suppressed the public edit logs for pages associated with the Russian invasion of Ukraine and encouraged editors to switch to alternate, less identifiable accounts as a means to protect their safety \cite{bri_wikipediawikipedia_2022, noauthor_wikipediawikipedia_2022}.

If users participate at all, they may only engage with taboo when they feel they have control over information about themselves: by interacting anonymously or developing a new or special-purpose account or profile \citep{taber_finsta_2020, menking_people_2019, kang_why_2013, forte_privacy_2017, birnholtz_is_2015}.
Because all these factors suggest the desirability of partially controlling or fully avoiding personal association with taboos, we propose \textbf{H5: peer produced resources about taboo subjects will be more likely to receive contributions from less identifiable personas than those about other comparable subjects.} 

\section{Empirical Setting}
\label{sec:setting}

Wikipedia is a vital source of information worldwide, with 318 language editions as of 2023.\footnote{\url{https://en.wikipedia.org/wiki/List_of_Wikipedias}} In December 2022 alone, Wikipedia received more than 24 million contributions and served up more than 23 billion page views across all projects, with 255,255 new user accounts created.\footnote{\url{https://stats.wikimedia.org}} The largest language edition, English, contains more than 7.5 million articles as of January 2023 and averages 559 new articles created each day.\footnote{\url{https://en.wikipedia.org/wiki/Wikipedia:Statistics}} Wikipedia entries are often the first result for a query on Google; text from relevant Wikipedia pages may be surfaced in search engine sidebars without the user needing to dig any further \citep{vincent_examining_2018}. 

\section{Methods}
\label{sec:methods}

\subsection{Technique for Identifying Taboo Subjects}

Our research questions concern taboo subjects in general. Therefore, we seek to minimize the assumptions we make about what is and is not taboo based on our own cultural context. 
Work in linguistics concerning taboo subjects has described how people use figurative language and euphemism to deal with taboo \citep{crespo_fernandez_conceptual_2011}. For example, we may say ``passed on'' instead of ``died.'' \citet{burridge_taboo_2015} describes euphemisms as ``a verbal escape hatch in response to taboos.'' %

To build our dataset of taboo articles in English Wikipedia, we made use of English Wiktionary.\footnote{\url{https://en.wiktionary.org/wiki/Wiktionary:Main_Page}} Wiktionary is a sister project to Wikipedia, building a peer produced dictionary. Wiktionary has received comparatively less attention from researchers when compared to the encyclopedia-building Wikipedia project.\footnote{For example, Wiktionary appears four times among the 1,972 examples of Wikipedia-related research listed in the Wikipedia page about Wikipedia research: \url{https://en.wikipedia.org/wiki/Wikipedia:List_of_academic_studies_about_Wikipedia}} Because Wiktionary entries are somewhat free-form, we used the \textit{wiktextract} Wiktionary parser and associated dataset to separate entries into key components for analysis: word, definition, and tags.\footnote{See \url{https://github.com/tatuylonen/wiktextract} and \url{https://kaikki.org/dictionary/}} In Wiktionary, definitions and tags are associated with each ``sense'' of a word to manage polysemy. Our corpus is composed of definitions remaining after we filtered the dataset to remove non-definitions (e.g., redirects to other words, definitions stating only that a word is a synonym or initialism of another word) and non-English definitions when the entry was supplying a translation. This filtration of 1,099,350 dictionary entries from the September 1, 2021 version of Wiktionary left us with 404,304 unique entries, each composed of a word, a definition, and, in some cases, associated tags.

\begin{figure}[t]
    \centering
    \includegraphics[width=5in]{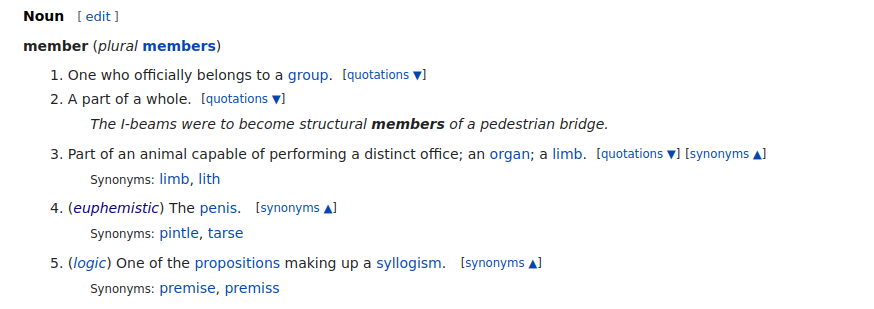}
    \caption{The Wiktionary definition of ``member''---which has meanings that range from group association to anatomical.}
    \label{fig:wiktEntry}
\end{figure}

Many dictionaries tag particular word usages as euphemistic. This makes dictionaries a source of general data on taboo because \textit{definitions} marked as euphemism contain a description of subjects considered taboo. For example, the term ``member'' has multiple definitions, one of which is marked as ``euphemistic'' in English Wiktionary (see Figure \ref{fig:wiktEntry}). The term being defined (i.e., ``member'') is not taboo---that is precisely why it is effective at creating distance from the taboo concept in the definition. 
We were inspired in this approach by \citet{buscaldi_sentiment_2015} who used the fact that Italian Wiktionary contributors have explicitly tagged some words as ``taboo'' to train a classifier for sentiment analysis. 

Our approach involved collecting only the text of definitions, not the words being defined. Tags in English Wiktionary can be applied to any definition. In Figure \ref{fig:wiktEntry}, ``(euphemistic)'' and ``(logic)'' are tags. We treated all definitions tagged as ``euphemistic'' as indicative of taboo. 
For example, in Figure \ref{fig:wiktEntry}, the definitions of ``member'' that refer to group association, a part of a whole, animal parts, and logic are each marked as not indicative of taboo in our dataset. However, the definition that refers to genitalia is tagged ``euphemistic'' and is marked as indicative of taboo in our dataset. 
We removed numbers and stop words, using the English stop words list in the Python Natural Language Tool Kit (NLTK), from all definitions.\footnote{\url{https://www.nltk.org/}} We also removed the words `term', `used', `usually', `particularly', `etc', `extremely', `especially', `one', `en', `something', `often', `synonym', `like', and `person' because these words recur very frequently in Wiktionary entries but convey only intensity. 

We used term frequency-inverse document frequency (TF-IDF) \citep{rajaraman_mining_2011} and ridge regression from the scikit-learn python package \citep{pedregosa_scikit-learn:_2011} 
to assess which words, bigrams, and trigrams (``n-grams'') were more commonly present in the definitions tagged as euphemistic. We sorted these terms by their coefficient from ridge regression to develop a list of 500 n-grams indicative of taboo. 

\begin{table}[t]
    \caption{A random selection of articles from our taboo set. We have omitted racial/ethnic slurs, explicit sexual acts, and profanity in this table but included them in our dataset.}
    \label{tab:tabooArticles}
    \centering
    \begin{tabular}{lllll}
    Sex differences in human physiology & Genitourinary system & Love God & Cat meat\\
    Gastrointestinal tract & Damnation & Alive and Dead & Old Age \\
    Monosodium glutamate & Being Bobby Brown & Abdominal Obesity & Feral cat \\
    Amazon Women on the Moon & Menstruation & Love and God  & 
    Layoff \\
    Mental disorder & Unemployment & Head on Collision & Vulva \\ 
    \end{tabular}
\end{table}

To test our hypotheses, it is not necessary that we detect all taboo articles in Wikipedia. Instead, we need only generate a dataset composed of two samples, one of which has a higher proportion of taboo articles.
In machine learning terms, we sought precision at the cost of recall. 
As a result, we performed an exact matching between Wikipedia article titles (minus stop words) and the 500 n-grams we identified as most associated with taboo. This conservative approach left us with a relatively small but high-confidence sample. Examples of these articles are listed in Table \ref{tab:tabooArticles}.\footnote{Inclusion of the article on the reality show ``Being Bobby Brown'' in the taboo set initially surprised us. The article is included because the article ``Hell to the no'' (a phrase the article says was popularized by the show) is a redirect that lands on this article. The words ``to'', ``the'' and ``no'' are stop words, leaving us with a match on the term ``Hell.''} 

For the comparison set, we considered using a sample of random Wikipedia articles. However, in exploratory analyses, we realized that this sample is inappropriate because most article titles in Wikipedia are not n-grams that appear in dictionary entries and could never have been selected by our taboo article identification process. Although many Wikipedia articles are phrases, events, locations, and proper names, the articles in our taboo set have a ``dictionary definition-esque'' quality. 
To ensure a comparable sample, we limit our comparison set to a random selection from the population of 115,681 articles in English Wikipedia with titles (minus stop words) that match an n-gram found in English Wiktionary definitions. 

\begin{figure}[t]
    \centering
    \includegraphics[width=.9\textwidth]{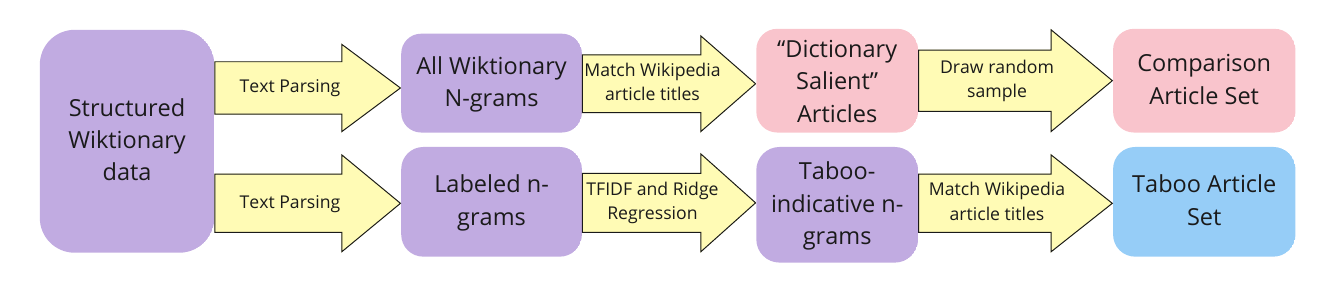}
    \caption{Our analytical pipeline first extracts n-grams, labeling them taboo if they are drawn from definitions tagged as euphemistic. Our samples are drawn from those articles that match these n-grams.} \label{fig:pipeline}
\end{figure}

We conducted additional processing on both the comparison set and the taboo set. Many Wikipedia pages serve only to disambiguate a term or provide a list of other articles. These non-articles were omitted. Further, some Wikipedia articles are ``redirects''---they serve as largely behind-the-scenes connectors to other articles and require special handling (see \citet{hill_consider_2014}.) When a term in our n-gram list was the title of an article serving as a redirect, we followed the redirect and extracted data for the target article. 
When the target of a redirect was a subsection of a longer page, 
we dropped the article from our sample because our measures cannot be easily constructed for article subsections. This left us with 74 taboo articles and 3,255 randomly selected comparison articles.

\subsubsection{Validation}
Evaluating our approach is difficult given the culturally varying nature of taboo and the lack of ground truth data. As a simple validity check, we test a hypothesis that
articles in our taboo set will be more likely to have been assigned Wikipedia categories related to sex---a subject that appears frequently in taboos across many cultural contexts \citep{moitra_understanding_2020,andalibi_understanding_2016,kannabiran_how_2011,crespo_fernandez_conceptual_2011,douglas_purity_1978,malinowski_crime_1926}---than those in our comparison set. %

Categories in Wikipedia are used for a wide range of purposes and are largely free form \cite{thornton_tagging_2012}. 
Past research by \citet{chen_effects_2010} and by \citet{asthana_few_2018} has shown the utility of categories associated with topically-focused WikiProjects \citep{morgan_project_2013} are a source of reliable information about an article's subject.
We used Wikimedia Foundation public APIs to obtain a list of 5,575 unique categories from the 3,329 articles in our datasets as well as their associated talk pages. 

We then used a logistic regression model to assess the relationship between sexual topics and membership in the taboo or comparison set in our sample. Articles can, and often do, exist in multiple categories simultaneously. 
As a baseline, taboo articles comprise 2.2\% of our sample. 
Given an article in our dataset that has been assigned to the scope of WikiProject Sexology and Sexuality, our model suggests that there is a 44\% chance of it being in the much smaller taboo set ($\beta= 3.4$, $z=8.43$, $p<0.001$).
This reflects very strong evidence that our taboo set is more likely to contain articles about sex and sexuality than is our comparison set.

\subsection{Data and Measures}

Having identified our taboo and comparison samples, we obtained the full text and metadata for revisions made to all articles in the samples by parsing the XML database dumps released by the Wikimedia Foundation.\footnote{\url{https://dumps.wikimedia.org}} 
Given that our interest is in human behavior, we attempt to omit revisions made by bots. To identify edits made by bots, we used a list we scraped from a Wikipedia page listing current registered bots as well as a historical dataset produced and released as part of \citet{geiger_operationalizing_2017}.
Our final article and revision datasets contained 177,974 revisions made to taboo articles and 2,052,209 revisions made to articles in the comparison set.
We used these data to construct longitudinal measures for each article in our sample: measures of quality of each article each month, the contribution history for each article, each contribution as well as its contributor, and the per-month viewership of the article. We describe each in turn.

In \textbf{H1} we suggest that taboo subjects will receive higher readership than comparable subjects. We operationalize readership using aggregate article view count data published by the Wikimedia Foundation to obtain the total number of views for each article in our sample for each month. We obtained 541,913 article-month measures of viewership and then calculated the mean view rank for each article.

To test \textbf{H2} on the number of contributions, we count the number of contributions made to each article in our sample over the life of each article.

In \textbf{H3}, we take up the question of whether taboo subjects receive lower-quality contributions. We operationalize contribution quality using two measures. The first, \textit{was reverted}, indicates that a given contribution was rejected after it was published. A contribution is said to have been reverted if someone restored an article to its state before the contribution in question. We looked 10 contributions forward to see if a contribution was reverted. 
The second measure we use, available through the ORES machine learning classifier's ``revision'' model, offers a prediction of whether or not a given contribution was ``damaging'' \citep{yang_identifying_2017}. In Wikipedia, damaging contributions are defined as those contributions that may need to be removed because they do not conform to Wikipedia guidelines. Damaging contributions can range from vandalism and misinformation to naive statements or formatting errors. Although many damaging contributions are simply reverted,
damaging contributions can instead be revised to be acceptable or be treated by community members as an opportunity to teach a new contributor \citep{halfaker_dont_2011}.

Our hypothesis \textbf{H4} concerns the quality of an article. We measure quality using the assessments generated using ORES' quality classifier \citep{halfaker_interpolating_2017}. The classifier was trained using the work of Wikipedia contributors and classifies revisions based on several structural features of articles, including length, the presence of pictures, and the use of links to other sources. %

In \textbf{H5}, we hypothesized that contributors to taboo subjects will be less identifiable. Because identifiability has many dimensions including the presence of unique personal identifiers, location, the presence or absence of a consistent pseudonym, and behavioral patterns \cite{marx_windows_2016}, we operationalize identifiability in several ways derived from prior work about identifiability on Wikipedia: (H5A) editing without an account \cite{forte_privacy_2017, anthony_reputation_2009}, (H5B) editing using a new account or one with fewer contributions \cite{forte_privacy_2017, rizoiu_evolution_2016, kraut_building_2012}, (H5C) revealing relatively little information on one's profile \cite{rizoiu_evolution_2016, bruckner_inferring_2021}, (H5D) identifying one's gender \cite{bruckner_inferring_2021}, and (H5E) setting one's account as ``emailable,'' which also requires a confirmed email address.

We measure whether a contribution was made with or without an account using metadata associated with each revision (H5A). We analyze contributor experience levels by counting the number of revisions each contributor has made to identify each revision as the contributor's ``$n$th edit'' (H5B). We use the XML database dumps to determine whether contributors had a user profile page at the time they were revising the articles in our samples (H5C), setting a dichotomous variable if they had a user page at any of these time points. Contributor gender and being emailable by others can be set by users in their Wikipedia setting page. We obtained this information via queries to the Wikimedia public API (H5D, H5E). 

Whether a contribution is made by a user with or without an account is highly visible to other Wikipedia users. Previous editing by a user is less visible but is still readily accessible. These signals have been previously found to influence how work is received in Wikipedia, with non-accountholders and accounts with lower editing counts being moderated more strictly \cite{teblunthuis_effects_2021}. 
These two identifying signals are also recorded automatically by the Wikipedia website software. The other signals are ``opt-in.'' User pages are free text profile pages and exist only if a user chooses to make one. Gender and emailability are only available via an API call, although emailability can be inferred by attempting to email the person via the website interface that is available to all logged-in users.

There is a potential confounder in our tests for H5A and H5B because some articles are ``protected'' so that they cannot be edited by contributors without accounts or by newcomers (those with accounts less than 4 days old and fewer than 10 edits).\footnote{\url{https://en.wikipedia.org/wiki/Wikipedia:Protection_policy}} 
In this way, page protection directly determines whether some anonymity seekers can participate. 

Following the method described in \citet{hill_page_2015}, we identify ``protection spells''---periods wherein a given page was protected. As \citet{hill_page_2015} describe, page protection data are left censored because of missing logs before 2008. As a result, we limit our assessment to the point at which reliable logs are available.
We use these data to
calculate the proportion of time each article in our dataset was protected since 2008. 
Articles varied in \textit{protected proportion} from 0 (e.g., the articles \textit{Abdominal Obesity} in our taboo set and \textit{Abbot} in our comparison set that were unprotected for the entire period) to more than 0.99 (e.g. the articles \textit{Hell} in our taboo set and \textit{Messianic Judaism} in our comparison set that were protected for almost the entire period). We use protected proportion as a control variable in our analyses for H5A and H5B.

\subsection{Ethics}
This study was conducted entirely using publicly available data published by the Wikimedia Foundation and does not involve any interaction or intervention with human subjects. This type of research using these data has been reviewed by the IRB at our institution and has been determined to not be human subject research. However, this work removes public digital trace data from its original context. Additionally, computational approaches have the potential to reveal behavioral trends in ways that individuals may find uncomfortable, especially given the subject of this study. As a result, we have redacted account names and IP addresses of the individuals who contributed to the articles in our sample. Article view data were fully anonymized by the Wikimedia Foundation prior to release.

Finally, we recognize that the use of taboo language may have differing impacts on those reading our work in various contexts. To make our work as available as possible while minimizing harm, we have redacted or omitted the use of racial/ethnic slurs and profanity in Table \ref{tab:tabooArticles} as well as in the text of our paper. However, all terms were included in the underlying analysis with no omissions. In the interest of accountability and open science, our full results and data are available in our dataset release and online supplement in an unabridged form. 

\section{Analytic Plan}
\label{sec:plan}

With respect to \textbf{H1} on viewership, our dataset includes articles written at different times and Wikipedia itself has received varying levels of traffic in the last two decades. Therefore, we calculated the rank in terms of views of each article with respect to other articles in our samples. We compared the mean within-month view rank of each article in our taboo and comparison samples across all months and tested for statistical significance using a Mann-Whitney U-test. The unit of analysis is the article, with mean rank across months as the outcome variable and taboo as our key predictor.

To test \textbf{H2} about contribution quantity, we examine the median number of contributions in our two samples and tested statistical significance using a Mann-Whitney U-test. The unit of analysis is the article, with total contributions as the outcome variable and taboo as our key predictor.

To test \textbf{H3} on contribution quality, we compared the median rate of reverted contributions of the two samples (total reverted/total contributions) and the median rate of damaging contributions, and tested statistical significance using a Mann-Whitney U-test. The unit of analysis is the article, with total reverted (damaging) contributions as the outcome variable and taboo as our key predictor.
As a robustness check, we conducted further analysis adding total contribution volume as a control variable in a linear regression.

To test \textbf{H4} about article quality, we compared the median quality of the two samples and tested for statistical significance using a Mann-Whitney U-test. The unit of analysis is the article, with our measure of average article quality as the outcome variable and taboo as our key predictor.

To evaluate \textbf{H5A} on the probability of someone contributing without an account varying between contributions to taboo and non-taboo articles, we fit a logistic regression on account status with taboo and a predictor and with protection as a control. The unit of analysis for this test is the revision. Because we have repeated measures of articles, we fit a multilevel model with a random intercept term for each article.
To evaluate \textbf{H5B} about the relationship of an article contribution being to a taboo article to contributor experience level, we fit a logistic regression on the logged average contributor experience with taboo as a predictor and page protection percentage as a control. The unit of analysis for this analysis is the article.

The identity affordances used as outcomes in tests for \textbf{H5C}, \textbf{H5D}, and \textbf{H5E} (user page, gender, emailability) are only available to accountholders, so we restrict these analyses to contributions made by accountholders. 
For H5C, H5D, and H5E, we used a user-level dataset.
Our analysis of H5C (user page) uses a dichotomous variable based on whether the contributor had a user page while making a contribution to any article in our sample. We use a chi-squared test to evaluate the relationship between having a user page and ever editing a taboo article.  
Our analysis of H5D (gender) uses two measures: first, someone must opt in to the gender-specifying feature, and second, they must choose a gender from the menu (the options given by the interface are ``male'' and ``female''). We tested these aspects separately. We evaluated the relationship between specifying a gender, specifying a gender that is female, and being emailable using a chi-squared test.

\section{Results}
\label{sec:results}

\subsection{Public Interest in Taboo Subjects (H1)}

\begin{figure}
    \centering
    \includegraphics[width=\textwidth]{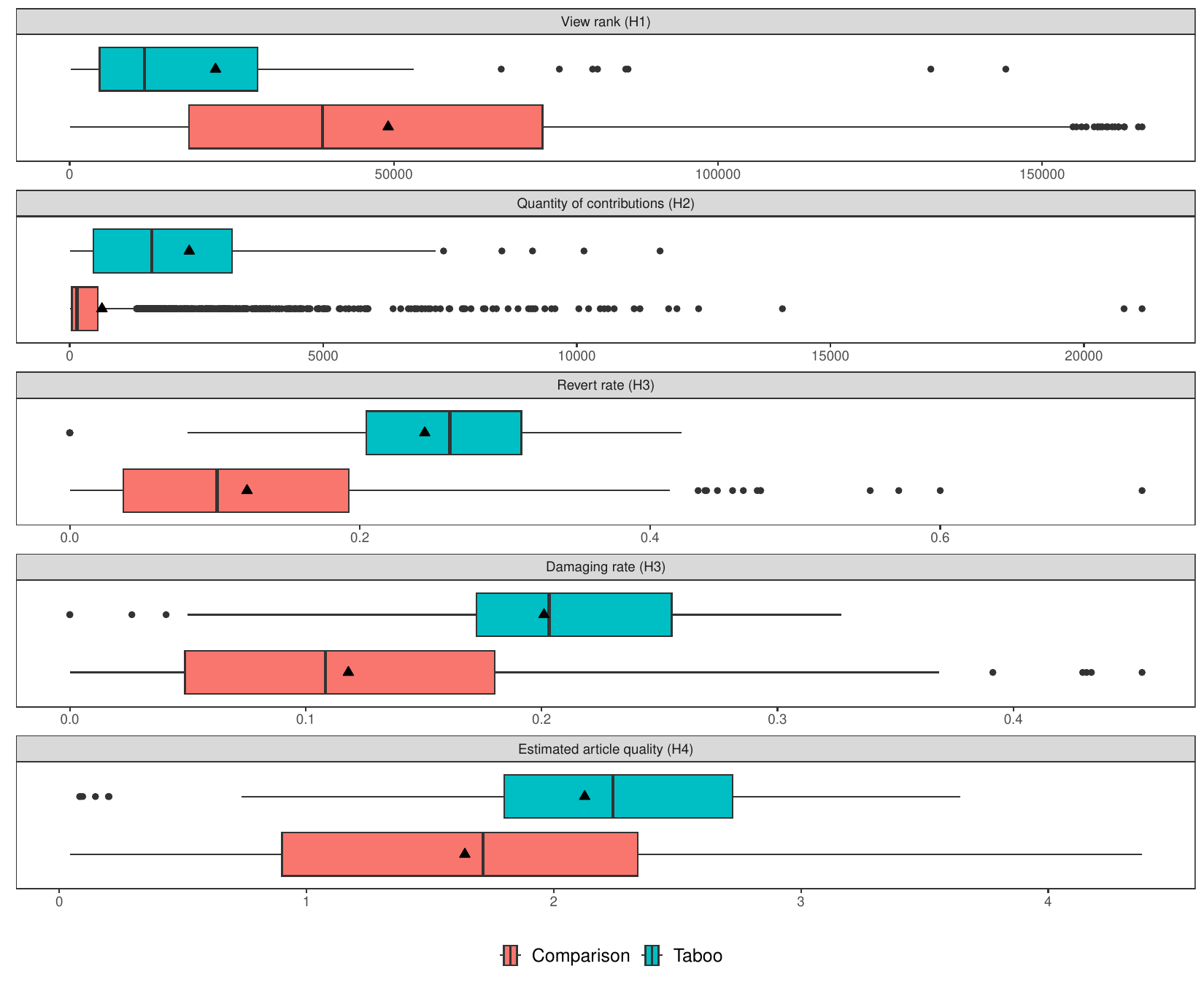}
    \caption{Boxplots showing the distributions of article-level variables for five hypothesis tests. From top to bottom: (a) view rank of articles where view rank is calculated within a given month across all articles where the most viewed article would rank 1 (H1); (b) quantity of contributions (H2); (c) article-level revert rates (H3); (d) article-level damaging contribution rate (H3); and (e) quality of the articles (H4).
    Small vertical lines in the boxes indicate medians. Triangles are located at the mean. }
    \label{fig:boxplots}
\end{figure}

Our analysis of page views provides evidence in support of H1 that taboo articles are viewed more frequently than those in our comparison set. The median view rank for taboo articles is 11,525 and the median view rank for the baseline comparison set is 39,017 ($U=189,375$, $p <0.001$). 
These data are represented in the boxplots in top panel of Figure \ref{fig:boxplots}. Note that the highest possible rank in this analysis is 1. We observe that although both sets have long tails including very unpopular articles, the higher popularity of taboo articles is such that the interquartile ranges of the two sets overlap only slightly. 

\subsection{Contribution Quantity (H2)}

Contrary to our expectations in \textbf{H2}, our analysis provides evidence that the quantity of contributions to taboo subjects is substantially higher than to the comparison set. The median number of contributions for taboo articles is 1,620, and the median number of contributions for the comparison set is 143 ($U=53,010$, $p<0.001$). These data are represented in the boxplots in the second panel of Figure \ref{fig:boxplots}.

\begin{table}[ht]

\begin{tabular}{l c}
\hline
 & Revert Count \\
\hline
(Intercept)        & $-27.3728^{*}$          \\
                   & $ [-30.6155; -24.1300]$ \\
Taboo              & $49.5379^{*}$           \\
                   & $ [ 29.9640;  69.1117]$ \\
Contribution Count & $0.2791^{*}$            \\
                   & $ [  0.2771;   0.2812]$ \\
\hline
R$^2$              & $0.9590$                \\
Adj. R$^2$         & $0.9590$                \\
Num. obs.          & $3299$                  \\
\hline
\multicolumn{2}{l}{\scriptsize{$^*$ 0 outside the confidence interval.}}
\end{tabular}

\caption{Linear regression model estimating effects for taboo on article revert count, with control for overall contribution count (H3).}  \label{tab:damageVolumeTab}
\end{table}

\subsection{Contribution Quality (H3)}

Our analysis of contribution quality provides support for  \textbf{H3}. We find that contributions to taboo articles are more frequently reverted than contributions to those in our comparison set. The median revert rate for taboo articles is 26.3\% and the median revert rate for the baseline comparison set is 10.1\% ($U=48,566$, $p<0.001$). These data are represented in the boxplots in the third panel of Figure \ref{fig:boxplots}.

We also measured contribution quality using the ORES model predicting ``damaging'' revisions and found that revisions to taboo articles are more frequently damaging than those to articles in our comparison set. The median damaging contribution rate for taboo articles is 20.3\% and the median damaging contribution rate for the baseline comparison set is 10.8\%, ($U=56,225$, $p<0.001$).
The distribution of the two sets are depicted in the fourth panel of Figure \ref{fig:boxplots}.
These estimates are similar in size to our estimates for our first hypothesis test for H3.

Both of these results for H3
provide evidence in support of our hypothesis that taboo subjects are the target of lower-quality contributions. However, given that our test of H2 showed that the number of revisions to taboo articles is an order of magnitude higher than the number of revisions to articles in our comparison set, edit volume appears to be a possible confounder. 
Indeed, revert rate and overall quantity of revisions are moderately positively correlated ($\rho=0.505$, $p <0.001$).
We report the results of our robustness check---a linear regression with a control for contribution count---in Table \ref{tab:damageVolumeTab}.
We observe that our prior finding that taboo articles receive more damaging contributions than do non-taboo articles is robust to the control for overall contribution volume.

\subsection{Article Quality (H4)}

Opposite to our proposal in \textbf{H4}, we find that articles addressing taboo subjects are higher quality than those in our comparison sample. The median quality level for taboo articles is 2.2 while the median quality for the comparison set is 1.7\ ($U=87,747$, $p<0.001$).
These data are visualized in the bottom panel of Figure \ref{fig:boxplots}.

\subsection{Contributor Identifiability (H5)}
In H5A-E, we proposed that contributors to taboo subjects would be less identifiable. 
The results for H5A about users with accounts are shown in Table \ref{tab:anonTab} and suggest that a revision to a taboo article is more likely to be made by a user without an account than a revision to the comparison set. The effect is as we hypothesized. Although the size of this effect is small (<1\%), a 0.05\% change in the 24 million contributions received in December 2022 is 120,000 contributions.

\begin{table}[ht]
\begin{tabular}{l c}
\hline
 & Contributing Without an Account \\
\hline
(Intercept)                   & $-0.9948^{*}$         \\
                              & $ [-1.0141; -0.9755]$ \\
Taboo                         & $0.1746^{*}$          \\
                              & $ [ 0.0547;  0.2944]$ \\
Protection Level              & $0.0791^{*}$          \\
                              & $ [ 0.0351;  0.1231]$ \\
\hline
AIC                           & $2749327$        \\
BIC                           & $2749377$        \\
Log Likelihood                & $-1374659$       \\
Num. obs.                     & $2219242$             \\
Num. groups: encodedTitle     & $3293$                \\
Var: encodedTitle (Intercept) & $0.2639$              \\
\hline
\multicolumn{2}{l}{\scriptsize{$^*$ 0 outside the confidence interval.}}
\end{tabular}

\caption{Hierarchical logistic regression model estimating the effects for taboo on whether or not a revision is made by someone contributing without an account (H5A). Fixed effects for the article are not shown.} 
\label{tab:anonTab}
\end{table}

In terms of H5B about editor experience, we find that the average editor to a taboo article has made fewer edits (21,621) than the average editor to our comparison set (36,176) ($U=171,617$; $p <0.001$). Further, in a model where we control for page protection (which blocks non-registered and low-experience contributors), contributors to taboo articles still tend to have fewer contributions. 
This result is surprising because one might expect that including this control would have reversed this relationship; blocking the least experienced contributors seems like it would be associated with contributors with more contributions rather than fewer, however this is not the case.
Page protection is also associated with fewer contributions, as expected. 
Table \ref{tab:linearExp} shows the results of a linear model with taboo as a predictor of (logged) contribution count and protection level as a control.

\begin{table}[t]
\begin{tabular}{l c}
\hline
 & Contribution Count Model \\
\hline
(Intercept)      & $10.6406^{*}$         \\
                 & $ [10.6122; 10.6689]$ \\
Taboo            & $-0.2318^{*}$         \\
                 & $ [-0.4143; -0.0492]$ \\
Protection Level & $-1.2622^{*}$         \\
                 & $ [-1.4679; -1.0565]$ \\
\hline
R$^2$            & $0.0477$              \\
Adj. R$^2$       & $0.0471$              \\
Num. obs.        & $3299$                \\
\hline
\multicolumn{2}{l}{\scriptsize{$^*$ 0 outside the confidence interval.}}
\end{tabular}
\caption{Results of a linear model examining the relationship between number of contributions a contributor has made (log scale) and whether they are contributing to a taboo subject with 95\% confidence intervals (H5B).}
\label{tab:linearExp}
\end{table}

Our analysis contradicts what we proposed in H5C. Of the 181,597 accountholders about whom we have user page data, 35.3\% of them had user pages at the point when they made any of the contributions in our sample. We found that 51.9\% of the contributors who ever contributed to a taboo article had a user page during one of their overall contributions to our samples, whereas only 32.8\% of those who did not contribute to a taboo article had user pages while contributing to a page in our sample, a statistically significant difference ($\chi^2=3267.2: p <0.001$).

These results provide evidence that accountholders who contributed to a taboo subject in our sample were more likely to have a user page than contributors in our sample who did not contribute to a taboo subject in our sample. 

Next, we observe that of the contributors who had ever contributed to the taboo articles in our sample, 17.7\% of them specified a gender, while of the contributors who did not contribute to the taboo articles, 10.0\% specified a gender ($\chi^2 = 1045.9: p <0.001$). This result contradicts \textbf{H5D} and suggests that people who contribute to taboo articles using an account are more likely to specify their gender. 
Out of curiosity, we also examined the relationship between taboo and the reported gender of those 19,678 individuals who did choose to report it. We found that 8.6\% of contributors to taboo articles who specify their gender specified female as compared to 8.9\%  in the comparison. We found that among account-holding contributors who choose to specify their gender, the relationship between reporting one's gender as female and contributing to a taboo subject is not statistically significant ($\chi^2 = 0.345: p = 0.557$).

Finally, we consider emailability. We find that 42.6\% of contributors to taboo articles have made themselves emailable, while 38.4\% of contributors to the comparison set have done so. We found that the relationship between editing a taboo article and setting oneself to be emailable is statistically significant ($\chi^2 = 149.8: p <0.001$). This result is contrary to \textbf{H5E}; contributors to taboo subjects are more likely to be emailable rather than less.

Of the five measures we used to operationalize different facets of identifiability, two were in the hypothesized direction: contributors to taboo subjects are less likely to have accounts and have less experience. We were surprised to find that accountholders who contribute to taboo subjects were more likely to have user pages, to reveal their gender and to make themselves emailable.

\section{Discussion}
\label{sec:discussion}

\subsection{The Success of Taboo Articles on Wikipedia}
Wikipedia articles on taboo subjects are very popular. Although one might suspect that societal opprobrium would suppress the quality of articles about taboo subjects, our results suggest the opposite is true. On Wikipedia, articles about taboo subjects receive more contributions and are higher quality than other similar articles. However, these articles also receive more low quality contributions than non-taboo articles do. %
We investigated whether techniques for limiting identifiability served as an important factor in the success of these articles and concluded that this relationship is not a simple one. Although contributors to taboo subjects were less likely to use accounts and tended to have less experience when compared to those who contribute to non-taboo subjects, we also found that that among the accountholding editors in our sample, those editing taboo subjects were more likely to have user pages, to specify their gender, and to make themselves emailable than those who did not edit taboo subjects. 

One potential explanation for our results with respect to identifiability may be the phenomena described in \citet{menking_people_2019} who described sophisticated means that women contributing to Wikipedia employ in order to contribute safely. Although \citet{menking_people_2019} describes ``choosing what to edit'' as a safety strategy, they also describe personal qualities that allow women to persist in the face of attacks. Contributors to taboo subjects may 
conceive of themselves as ``people who can take it'' in the words of one of  \citepos{menking_people_2019} interviewees.

Or, in that the success of peer production projects may be tied to the diversity of motives from participants \citep{benkler_wealth_2006}, it may be that contributors to taboo subjects are specifically motivated with respect to the taboo subjects to which they choose to contribute. For example, they may take an advocacy position with respect to women's health information. Investigating this subject further might require additional methods such as participant interviews. Of course, an interview approach may struggle to include those who contribute without accounts and more casual contributors who may be harder to reach and recruit for interviews. 

Taken as a whole, our findings introduce an empirical puzzle: How is it that taboo articles are higher quality than non-taboo articles 
despite the larger number and rate of poor quality and damaging contributions that these articles receive from less identifiable contributors? One possible explanation may be a mechanism \citet{gorbatai_paradox_2014} elaborates: perhaps these unhelpful contributions serve as a signal of public interest in a subject and draw in experienced editors to clean up the mess left by less experienced contributors and who also, perhaps, further improve the article while doing so. 
Our result provides further evidence that the end result of unhelpful contributions may depend on how the community responds, and reinforces the observation from \citet{hill_hidden_2021} that increasing barriers to novice contributions (which may indeed be lower quality) may have deleterious effects.

\subsection{Further Exploration of Article Quality}
\begin{figure}[t]
\includegraphics[width=0.7\textwidth]{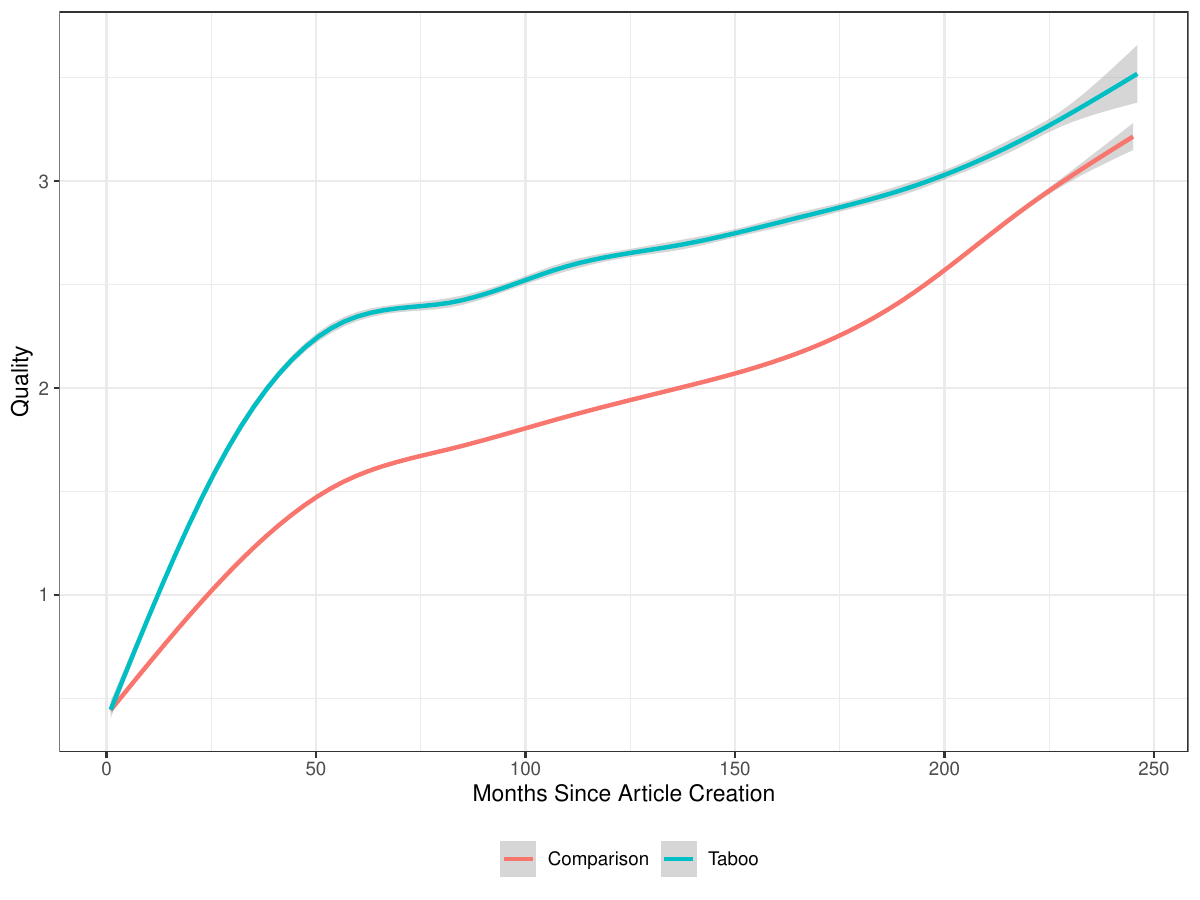}
\caption{Visualization of average article quality over time as predicted by the Wikimedia ORES API shown using generalized additive model (GAM) smoothers. We see that in the first several years of their existence, taboo subjects grow somewhat more quickly in quality, but that their quality growth over time begins to track more closely to the comparison set.}
\label{fig:ageGrowth}
\end{figure}  

\begin{figure}[t]
\includegraphics[width=0.7\textwidth]{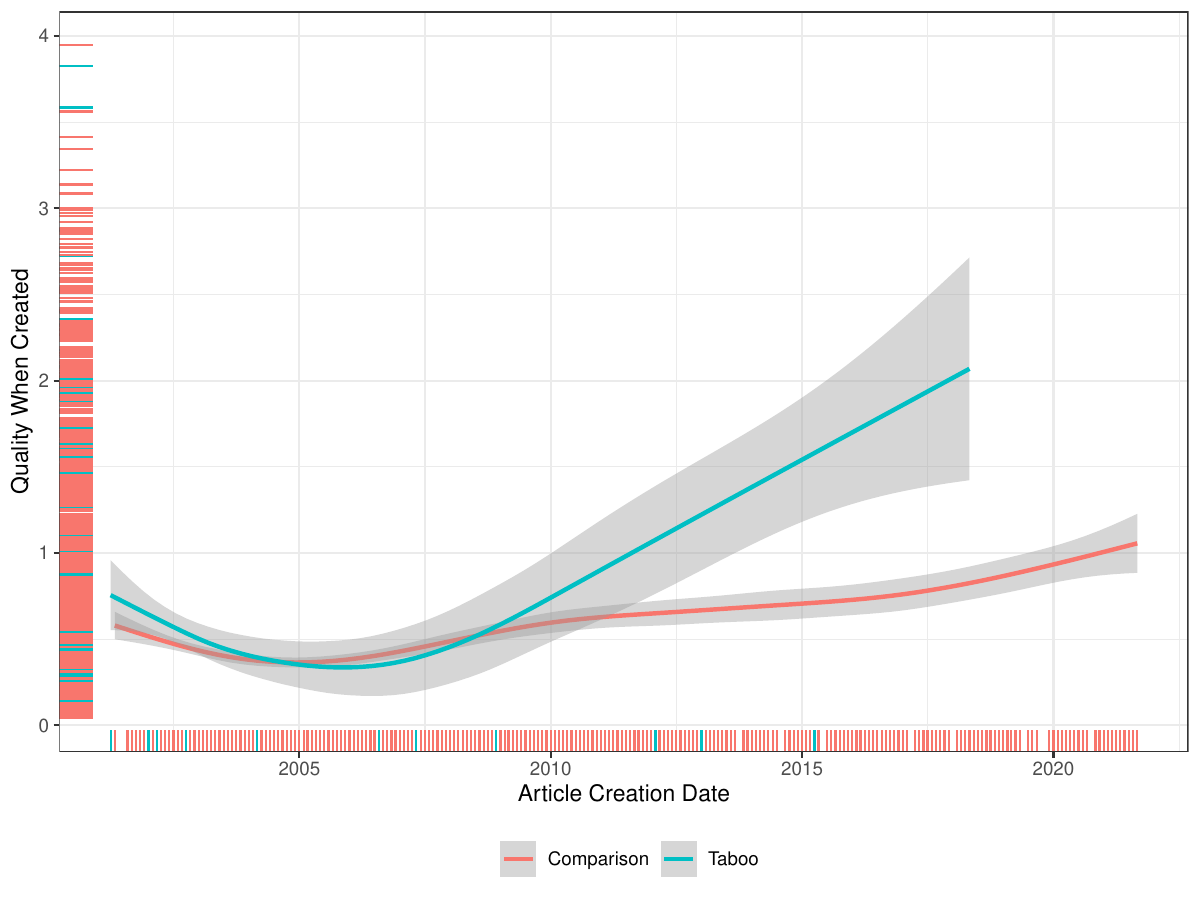}
\caption{Average quality of the first version of new articles over time shown using generalized additive model (GAM) smoothers. The rug along the axes identifies the areas with the greatest concentration of data.} \label{fig:qualityAtBirth}
\end{figure}  

Given our surprising finding on the quality of articles about taboo subjects, we conducted additional analysis. Figure \ref{fig:ageGrowth} visualizes the quality growth trajectory of taboo and comparison articles over time using local regression (LOESS) models. Quality is as predicted by the Wikimedia ORES model on a per-article-per-month level. 
Articles in these two sets are different early in their existence but generally follow similar trajectories after several years. 
Figure \ref{fig:qualityAtBirth} shows the average quality of articles when they are created. 
The initial quality of taboo articles and comparison articles in the early period of Wikipedia were fairly similar, but recently created articles are higher quality at the point of creation. 
Additionally, we have included a cross-tabulation of whether contributors were accountholders, which sample they appear in, and whether their contribution was reverted, in our online supplement.
These results suggest that one potential explanation for the high quality of taboo articles is survival bias.
For an article to survive long enough to enter our dataset, especially those articles created recently, it must have had sufficient quality when created in order to convey its value. 
Initial quality and growth place taboo articles further ahead in their development, even as maintenance and enhancement of both groups of articles eventually follow similar trajectories.  
One explanation for the preliminary differences we observe may be that individual editors exercising quality control in Wikipedia may scrutinize new taboo articles even more closely than they do other articles, perhaps drawn in by their own interest in taboo subjects or a sense that taboo subjects may draw in the creation of spurious articles. 

Another part of the explanation for the quality of taboo articles may be the presence of health-related content in the taboo set. Wikipedia has specialized guidelines around health and medicine sourcing and is home to an organized effort to uphold those guidelines (WikiProject Medicine) \citep{james_wikiproject_2016,trevena_wikiproject_2011}. However, this is a question in need of additional research because many such organizations and content guidelines exist in Wikipedia, and because the existence of strict guidelines and the presence of a group enforcing those guidelines could serve to decrease the proportion of contributions ultimately deemed acceptable.

\subsection{Implications For Design}

Our results indicate that there is substantial reason for optimism with respect to the availability of knowledge about taboo subjects and that online communities like Wikipedia can, in some cases, resist societal norms in ways that support human thriving. 
This result is particularly striking because our setting is Wikipedia. As a tertiary source, an encyclopedia is dependent on the availability of secondary content in the form of publications from reliable sources. There may be much more we can learn from this success as we seek to replicate it in other settings.

Contributors to taboo subjects contradicted our expectations of low quality and maximal privacy. 
Our finding that contributors to taboo subjects used some privacy affordances but not others suggests that collaborative platforms should be wary of assuming that participants have a uniform perspective on their privacy. Indeed some may prefer to make themselves more identifiable despite the potential increased risk. However, policy changes that would increase requirements for identifiability may endanger this success, at least with respect to the contributors who appear in our data as non-accountholders or who have a lower overall edit count. Many peer production projects, including some language editions of Wikipedia, have policies requiring that participants make themselves even more identifiable than they are required to be on English Wikipedia. Our finding that contributors take a variable approach to privacy suggests designers exercise caution when considering privacy-sacrificing features because a lack of privacy may limit the participation of people seeking to engage with taboo subjects. Further, these results suggest that future studies of taboo in social computing systems should examine privacy-seeking behaviors among different roles (e.g., reader, contributor, administrator). 

We also observe that identifiability is multifaceted. In this case, we find differences even in these relatively weak identifiers: these are not ``real name'' policies, user pages could contain content of any kind, none of the information being provided is verified, and although participants can opt into emailability, doing so does not expose their email address. 
Affordances that seek to make participants more identifiable on platforms have been found to have unwanted side effects that go beyond discouraging participation \citep{hill_hidden_2021}. Additionally, diminishing contributor choices on what information about them is collected and shared may diminish their willingness to contribute to subjects that may place them at risk. In turn, this could lead to lower-quality, neglected articles. Such an outcome would disproportionately harm information-seekers most in need of reliable information.

\subsection{Implications for Future Research}
Our methods have potential utility for future studies of taboo across HCI and across languages. Although we selected English corpora, Wiktionary exists in 186 different languages and Wikipedia in more than 300. Further, although there may be some advantage to the fact that both Wiktionary and Wikipedia operate under the same umbrella foundation and have overlapping communities, other dictionaries and sites of collaborative production could be examined in a similar manner. 

Numerous other dictionaries use sense tagging to indicate that a definition corresponds to euphemistic usage. Further, there are many other sense tags in use, including the identification of slang, expletives, epithets, jargon, dialect, and archaic usage. Connecting linguistic scholarship about these forms of language to other social computing phenomena may be fruitful.  Further, a computational approach of this kind is scalable across multiple settings, including those where human inspection is infeasible because of volume. We have made an anonymized dataset and our analytical code available for replication and reuse of our findings at https://doi.org/10.7910/DVN/5OKEEO.

\section{Limitations}
\label{sec:limitations}
Our method of taboo identification through NLP techniques lacks important nuance and context. Because taboo varies across cultures, behaviors that are profoundly polluting for some may carry little or no taboo for others. We are researchers with relatively high levels of privilege with respect to freedom of inquiry; this positionality has inevitably influenced our results. Our profession, discipline, and employers act to legitimize our engagement with taboo subjects, which may in turn desensitize us to the risks of violating taboo.

Our ability to generalize these results to other contexts is limited by the fact that
we only examined English language Wikipedia. This limitation reflects our  language knowledge, given the deeply linguistic nature of the subject matter.
Given that taboo is substantially influenced by local and cultural factors, our findings might be different in other language contexts. We hope that authors with a broader range of language backgrounds will take up similar questions in the future.

Additionally, contributions to Wiktionary, from which we drew our original list of key words indicating taboo, may be biased. We know that Wikipedia suffers from numerous participation gaps with respect to gender, skills, and race and we imagine these gaps might extend to Wiktionary as well in ways that might shape the kinds of taboo that is reflected in our dataset \citep{shaw_pipeline_2018, hargittai_online_2016, hargittai_mind_2015, hill_wikipedia_2013}. Although we were unable to find data on participation gaps in Wiktionary, surveys conducted across all Wikimedia Foundation projects (which include Wikipedia and Wiktionary) have found gaps in participation.\footnote{\url{https://meta.wikimedia.org/wiki/Community_Insights/Community_Insights_2021_Report}} Given that taboo functions as social control, insofar as similar contributor demographic gaps are present in Wiktionary, our identification of taboo subjects likely reflects the attitudes of the people who are more privileged and have more means to resist the consequences of taboo violations. Although this remains an important unaddressed threat to validity, we believe that it likely makes our method of identifying taboo conservative, if potentially biased.

Further, Wiktionary has limitations as a source of data on taboo in that its self-organized volunteer contributor nature and openness to non-experts may lead to inconsistencies; euphemistic definitions may go untagged while non-euphemisms may be erroneously tagged as such. For example, our taboo dataset includes the article \textit{Super Bowl}. The bigram ``super bowl'' is in our taboo n-grams list because out of the five times it appears in definitions in Wiktionary, two of them (the terms ``superb owl'' and ``Big Game'') have definitions that are tagged as euphemism. Although we would argue that this is not a correct usage of the euphemism tag (instead, these terms seem more like slang), we have included the Super Bowl article out of fidelity to the source material. Taking on this descriptivist approach can add noise, but can also expand our perspective. When we observed that our dataset included euphemisms around nuclear weapons and napalm, we realized that these topics were not what we had been expecting when we first began to explore taboo. Indeed, they might have escaped our notice. However, once sensitized to this taboo through our systematic approach, we were able to validate it and found that prior work has indeed explored the taboo surrounding nuclear and chemical weapons \citep{schelling_nuclear_2007, price_chemical_2007, cole_poison_1998}.

Our measures are limited as well. Views data only examines page hits---not whether the article was read. Contributions vary in their size, and their assessed quality is subject to community values and biases. The measure of article quality we have used, although common in Wikipedia research, only assesses articles based on observable features such as length, links, and the presence of images: it does not gauge content, and collapses quality down to a single continuous measure despite being derived from an ordinal system of quality classes that are not evenly spaced \cite{teblunthuis_measuring_2021}.  

We used several measures to assess contributor identifiability. By constructing a user-level dataset with dichotomous variables for H5C, D, and E, we omit variation over time, and we do not consider the inequalities in participation between participants; some descriptive analysis of a revision-level dataset is in our online supplement. Although we use the presence of profiles as a sign of identifiability, people tell ``privacy lies'' by entering false information into online profiles---sometimes in order to to avoid being stigmatized \citep{sannon_privacy_2018}. We make no effort to address whether user pages contain identifying information or whether the information provided is correct. Selecting a gender from a drop-down menu may or may not align with aspects of someone's gender identity. Additionally, although we treat contributing without an account as a sign of lower identifiability, the IP addresses made visible as part of that contribution can be highly identifying. %
Finally, we only have information on whether these traits were present at the time we made our API queries in June 2022, and do not have historical information as to whether gender or emailability had been set at the moment of contribution. 
We attempt to mitigate these limitations through our use of several different measures to assess identifiability. 

Finally, although our method seeks to minimize our assumptions about what is taboo, we made many choices throughout the process that may have influenced our results. We chose stop words. We relied on our own sense of what qualifies as taboo when checking our results for face validity during all points of conducting this research project. Although we believe that these ongoing informal assessments were consistent with the work of prior researchers, our own perspective inevitably entered into the conduct of this work. 

\section{Conclusion}
\label{sec:conclusion}
Information-seekers turn to social computing systems to learn about subjects that society has gauged to be taboo, and Wikipedia contributors have successfully built resources to serve this need.  
This work makes three contributions: (1) we elaborate a computational technique for detecting taboo based on euphemistic dictionary definitions, (2) we illustrate this approach by applying it to English Wikipedia, and (3) we analyze distinctions in both how taboo subjects develop and who contributes to the development of information about taboo subjects. Future work should explore the relationships between identifiability and taboo subjects in other environments, continue to unpack the social dynamics of taboo, and consider the impacts of policy changes that diminish access to privacy-preserving technologies.
Taboo subjects connect to some of the most fundamental pieces of human existence and both some of the most distressing and most uplifting parts of our social experience. 
Although we as researchers may recoil from doing work on ``not safe for work'' subjects, facing taboos and tackling them as area of inquiry is worth the discomfort it may give us. Taboos shape our consumption and production behaviors across social computing systems. Exploring taboo subjects through the lens of HCI is a vital and tranformative area of future work for our field.

\begin{acks}
This work benefited substantively from the guidance of anonymous reviewers and the program committee at CSCW. The creation of our dataset was aided by the use of advanced computational, storage, and networking infrastructure provided by the Hyak supercomputer system at the University of Washington and high-performance computing systems supported by the Northwestern School of Communication and the National Science Foundation. This work was supported by the National Science Foundation (awards CNS-1703736 and CNS-1703049).
\end{acks}

\section{Data and Code}
A replication dataset including data, code, and other supplementary material has been placed in the Harvard Dataverse archive and is available at: https://doi.org/10.7910/DVN/5OKEEO

\bibliographystyle{ACM-Reference-Format}
\bibliography{refs}


\begin{thebibliography}{90}


\ifx \showCODEN    \undefined \def \showCODEN     #1{\unskip}     \fi
\ifx \showDOI      \undefined \def \showDOI       #1{#1}\fi
\ifx \showISBNx    \undefined \def \showISBNx     #1{\unskip}     \fi
\ifx \showISBNxiii \undefined \def \showISBNxiii  #1{\unskip}     \fi
\ifx \showISSN     \undefined \def \showISSN      #1{\unskip}     \fi
\ifx \showLCCN     \undefined \def \showLCCN      #1{\unskip}     \fi
\ifx \shownote     \undefined \def \shownote      #1{#1}          \fi
\ifx \showarticletitle \undefined \def \showarticletitle #1{#1}   \fi
\ifx \showURL      \undefined \def \showURL       {\relax}        \fi
\providecommand\bibfield[2]{#2}
\providecommand\bibinfo[2]{#2}
\providecommand\natexlab[1]{#1}
\providecommand\showeprint[2][]{arXiv:#2}

\bibitem[\protect\citeauthoryear{??}{noa}{2022}]%
        {noauthor_wikipediawikipedia_2022}
 \bibinfo{year}{2022}\natexlab{}.
\newblock \bibinfo{title}{Wikipedia:{Wikipedia} {Signpost}/2022-04-24/{In}
  focus}.
\newblock
\newblock
\urldef\tempurl%
\url{https://en.wikipedia.org/w/index.php?title=Wikipedia:Wikipedia_Signpost/2022-04-24/In_focus&oldid=1090495808}
\showURL{%
\tempurl}


\bibitem[\protect\citeauthoryear{Allan}{Allan}{2019}]%
        {allan_taboo_2019}
\bibfield{author}{\bibinfo{person}{Keith Allan}.}
  \bibinfo{year}{2019}\natexlab{}.
\newblock \showarticletitle{Taboo words and language: {An} overview}.
\newblock In \bibinfo{booktitle}{\emph{The {Oxford} handbook of taboo words and
  language} (\bibinfo{edition}{first edition} ed.)},
  \bibfield{editor}{\bibinfo{person}{Keith Allan}} (Ed.).
  \bibinfo{publisher}{Oxford University Press}, \bibinfo{address}{Oxford ; New
  York, NY}.
\newblock
\showISBNx{978-0-19-880819-0}


\bibitem[\protect\citeauthoryear{Almeida, Comber, Wood, Saraf, and
  Balaam}{Almeida et~al\mbox{.}}{2016}]%
        {almeida_looking_2016}
\bibfield{author}{\bibinfo{person}{Teresa Almeida}, \bibinfo{person}{Rob
  Comber}, \bibinfo{person}{Gavin Wood}, \bibinfo{person}{Dean Saraf}, {and}
  \bibinfo{person}{Madeline Balaam}.} \bibinfo{year}{2016}\natexlab{}.
\newblock \showarticletitle{On {Looking} at the {Vagina} through {Labella}}. In
  \bibinfo{booktitle}{\emph{Proceedings of the 2016 {CHI} {Conference} on
  {Human} {Factors} in {Computing} {Systems}}}. \bibinfo{publisher}{ACM},
  \bibinfo{address}{San Jose California USA}, \bibinfo{pages}{1810--1821}.
\newblock
\showISBNx{978-1-4503-3362-7}
\urldef\tempurl%
\url{https://doi.org/10.1145/2858036.2858119}
\showDOI{\tempurl}


\bibitem[\protect\citeauthoryear{Ammari, Schoenebeck, and Romero}{Ammari
  et~al\mbox{.}}{2019}]%
        {ammari_self-declared_2019}
\bibfield{author}{\bibinfo{person}{Tawfiq Ammari}, \bibinfo{person}{Sarita
  Schoenebeck}, {and} \bibinfo{person}{Daniel Romero}.}
  \bibinfo{year}{2019}\natexlab{}.
\newblock \showarticletitle{Self-declared throwaway accounts on reddit: how
  platform affordances and shared norms enable parenting disclosure and
  support}.
\newblock \bibinfo{journal}{\emph{Proceedings of the ACM on Human-Computer
  Interaction}} \bibinfo{volume}{3}, \bibinfo{number}{CSCW}
  (\bibinfo{date}{Nov.} \bibinfo{year}{2019}), \bibinfo{pages}{135:1--135:30}.
\newblock
\urldef\tempurl%
\url{https://doi.org/10.1145/3359237}
\showDOI{\tempurl}


\bibitem[\protect\citeauthoryear{Andalibi}{Andalibi}{2020}]%
        {andalibi_disclosure_2020}
\bibfield{author}{\bibinfo{person}{Nazanin Andalibi}.}
  \bibinfo{year}{2020}\natexlab{}.
\newblock \showarticletitle{Disclosure, {Privacy}, and {Stigma} on {Social}
  {Media}: {Examining} {Non}-{Disclosure} of {Distressing} {Experiences}}.
\newblock \bibinfo{journal}{\emph{ACM Transactions on Computer-Human
  Interaction}} \bibinfo{volume}{27}, \bibinfo{number}{3} (\bibinfo{date}{June}
  \bibinfo{year}{2020}), \bibinfo{pages}{1--43}.
\newblock
\showISSN{1073-0516, 1557-7325}
\urldef\tempurl%
\url{https://doi.org/10.1145/3386600}
\showDOI{\tempurl}


\bibitem[\protect\citeauthoryear{Andalibi, Haimson, De~Choudhury, and
  Forte}{Andalibi et~al\mbox{.}}{2016}]%
        {andalibi_understanding_2016}
\bibfield{author}{\bibinfo{person}{Nazanin Andalibi},
  \bibinfo{person}{Oliver~L. Haimson}, \bibinfo{person}{Munmun De~Choudhury},
  {and} \bibinfo{person}{Andrea Forte}.} \bibinfo{year}{2016}\natexlab{}.
\newblock \showarticletitle{Understanding {Social} {Media} {Disclosures} of
  {Sexual} {Abuse} {Through} the {Lenses} of {Support} {Seeking} and
  {Anonymity}}. In \bibinfo{booktitle}{\emph{Proceedings of the 2016 {CHI}
  {Conference} on {Human} {Factors} in {Computing} {Systems}}}.
  \bibinfo{publisher}{ACM}, \bibinfo{address}{San Jose California USA},
  \bibinfo{pages}{3906--3918}.
\newblock
\showISBNx{978-1-4503-3362-7}
\urldef\tempurl%
\url{https://doi.org/10.1145/2858036.2858096}
\showDOI{\tempurl}


\bibitem[\protect\citeauthoryear{Anthony, Smith, and Williamson}{Anthony
  et~al\mbox{.}}{2009}]%
        {anthony_reputation_2009}
\bibfield{author}{\bibinfo{person}{Denise~L. Anthony}, \bibinfo{person}{Sean~W.
  Smith}, {and} \bibinfo{person}{Timothy Williamson}.}
  \bibinfo{year}{2009}\natexlab{}.
\newblock \showarticletitle{Reputation and reliability in collective goods:
  {The} case of the online encyclopedia {Wikipedia}}.
\newblock \bibinfo{journal}{\emph{Rationality and Society}}
  \bibinfo{volume}{21}, \bibinfo{number}{3} (\bibinfo{date}{Aug.}
  \bibinfo{year}{2009}), \bibinfo{pages}{283--306}.
\newblock
\showISSN{1043-4631}
\urldef\tempurl%
\url{https://doi.org/10.1177/1043463109336804}
\showDOI{\tempurl}


\bibitem[\protect\citeauthoryear{Asthana and Halfaker}{Asthana and
  Halfaker}{2018}]%
        {asthana_few_2018}
\bibfield{author}{\bibinfo{person}{Sumit Asthana} {and} \bibinfo{person}{Aaron
  Halfaker}.} \bibinfo{year}{2018}\natexlab{}.
\newblock \showarticletitle{With {Few} {Eyes}, {All} {Hoaxes} {Are} {Deep}}.
\newblock \bibinfo{journal}{\emph{Proc. ACM Hum.-Comput. Interact.}}
  \bibinfo{volume}{2}, \bibinfo{number}{CSCW} (\bibinfo{date}{Nov.}
  \bibinfo{year}{2018}), \bibinfo{pages}{21:1--21:18}.
\newblock
\showISSN{2573-0142}
\urldef\tempurl%
\url{https://doi.org/10.1145/3274290}
\showDOI{\tempurl}


\bibitem[\protect\citeauthoryear{Aucoin}{Aucoin}{2021}]%
        {aucoin_censorship_2021}
\bibfield{author}{\bibinfo{person}{Jessica Aucoin}.}
  \bibinfo{year}{2021}\natexlab{}.
\newblock \showarticletitle{Censorship in {Libraries}: {A} {Retrospective}
  {Study} of {Banned} and {Challenged} {Books}}.
\newblock \bibinfo{journal}{\emph{SLIS Connecting}} \bibinfo{volume}{10},
  \bibinfo{number}{2} (\bibinfo{year}{2021}), \bibinfo{pages}{82--100}.
\newblock
\showISSN{23302917}
\urldef\tempurl%
\url{https://doi.org/10.18785/slis.1002.09}
\showDOI{\tempurl}


\bibitem[\protect\citeauthoryear{Benkler}{Benkler}{2002}]%
        {benkler_coases_2002}
\bibfield{author}{\bibinfo{person}{Yochai Benkler}.}
  \bibinfo{year}{2002}\natexlab{}.
\newblock \showarticletitle{Coase's {Penguin}, or, {Linux} and "{The} {Nature}
  of the {Firm}"}.
\newblock \bibinfo{journal}{\emph{The Yale Law Journal}} \bibinfo{volume}{112},
  \bibinfo{number}{3} (\bibinfo{date}{Dec.} \bibinfo{year}{2002}),
  \bibinfo{pages}{369}.
\newblock
\showISSN{00440094}
\urldef\tempurl%
\url{https://doi.org/10.2307/1562247}
\showDOI{\tempurl}


\bibitem[\protect\citeauthoryear{Benkler}{Benkler}{2006}]%
        {benkler_wealth_2006}
\bibfield{author}{\bibinfo{person}{Yochai Benkler}.}
  \bibinfo{year}{2006}\natexlab{}.
\newblock \bibinfo{booktitle}{\emph{The wealth of networks: {How} social
  production transforms markets and freedom}}.
\newblock \bibinfo{publisher}{Yale University Press}, \bibinfo{address}{New
  Haven, CT}.
\newblock


\bibitem[\protect\citeauthoryear{Birnholtz, Merola, and Paul}{Birnholtz
  et~al\mbox{.}}{2015}]%
        {birnholtz_is_2015}
\bibfield{author}{\bibinfo{person}{Jeremy Birnholtz}, \bibinfo{person}{Nicholas
  Aaron~Ross Merola}, {and} \bibinfo{person}{Arindam Paul}.}
  \bibinfo{year}{2015}\natexlab{}.
\newblock \showarticletitle{"{Is} it {Weird} to {Still} {Be} a {Virgin}":
  {Anonymous}, {Locally} {Targeted} {Questions} on {Facebook} {Confession}
  {Boards}}. In \bibinfo{booktitle}{\emph{Proceedings of the 33rd {Annual}
  {ACM} {Conference} on {Human} {Factors} in {Computing} {Systems}}}.
  \bibinfo{publisher}{ACM}, \bibinfo{address}{Seoul Republic of Korea},
  \bibinfo{pages}{2613--2622}.
\newblock
\showISBNx{978-1-4503-3145-6}
\urldef\tempurl%
\url{https://doi.org/10.1145/2702123.2702410}
\showDOI{\tempurl}


\bibitem[\protect\citeauthoryear{Bri and EpicPupper}{Bri and
  EpicPupper}{2022}]%
        {bri_wikipediawikipedia_2022}
\bibfield{author}{\bibinfo{person}{Bri} {and} \bibinfo{person}{EpicPupper}.}
  \bibinfo{year}{2022}\natexlab{}.
\newblock \bibinfo{title}{Wikipedia:{Wikipedia} {Signpost}/2022-03-27/{News}
  and notes}.
\newblock
\newblock
\urldef\tempurl%
\url{https://en.wikipedia.org/w/index.php?title=Wikipedia:Wikipedia_Signpost/2022-03-27/News_and_notes&oldid=1084492254}
\showURL{%
\tempurl}
\newblock
\shownote{Page Version ID: 1084492254.}


\bibitem[\protect\citeauthoryear{Brown}{Brown}{2016}]%
        {brown_female_2016}
\bibfield{author}{\bibinfo{person}{Taylor~Kate Brown}.}
  \bibinfo{year}{2016}\natexlab{}.
\newblock \showarticletitle{Female scientist fights harassment with
  {Wikipedia}}.
\newblock \bibinfo{journal}{\emph{BBC News}} (\bibinfo{date}{March}
  \bibinfo{year}{2016}).
\newblock
\urldef\tempurl%
\url{https://www.bbc.com/news/blogs-trending-35787730}
\showURL{%
\tempurl}


\bibitem[\protect\citeauthoryear{Brückner, Lemmerich, and
  Strohmaier}{Brückner et~al\mbox{.}}{2021}]%
        {bruckner_inferring_2021}
\bibfield{author}{\bibinfo{person}{Sebastian Brückner},
  \bibinfo{person}{Florian Lemmerich}, {and} \bibinfo{person}{Markus
  Strohmaier}.} \bibinfo{year}{2021}\natexlab{}.
\newblock \showarticletitle{Inferring {Sociodemographic} {Attributes} of
  {Wikipedia} {Editors}: {State}-of-the-art and {Implications} for {Editor}
  {Privacy}}. In \bibinfo{booktitle}{\emph{Companion {Proceedings} of the {Web}
  {Conference} 2021}}. \bibinfo{publisher}{ACM}, \bibinfo{address}{Ljubljana
  Slovenia}, \bibinfo{pages}{616--622}.
\newblock
\showISBNx{978-1-4503-8313-4}
\urldef\tempurl%
\url{https://doi.org/10.1145/3442442.3452350}
\showDOI{\tempurl}


\bibitem[\protect\citeauthoryear{Burridge}{Burridge}{2015}]%
        {burridge_taboo_2015}
\bibfield{author}{\bibinfo{person}{Kate Burridge}.}
  \bibinfo{year}{2015}\natexlab{}.
\newblock \showarticletitle{Taboo {Words}}.
\newblock In \bibinfo{booktitle}{\emph{The {Oxford} {Handbook} of the {Word}}},
  \bibfield{editor}{\bibinfo{person}{John~R. Taylor}} (Ed.).
\newblock
\showISBNx{978-0-19-964160-4}
\urldef\tempurl%
\url{https://doi.org/10.1093/oxfordhb/9780199641604.013.017}
\showDOI{\tempurl}


\bibitem[\protect\citeauthoryear{Buscaldi and Hernandez-Farias}{Buscaldi and
  Hernandez-Farias}{2015}]%
        {buscaldi_sentiment_2015}
\bibfield{author}{\bibinfo{person}{Davide Buscaldi} {and}
  \bibinfo{person}{Irazú Hernandez-Farias}.} \bibinfo{year}{2015}\natexlab{}.
\newblock \showarticletitle{Sentiment {Analysis} on {Microblogs} for {Natural}
  {Disasters} {Management}: a {Study} on the 2014 {Genoa} {Floodings}}. In
  \bibinfo{booktitle}{\emph{Proceedings of the 24th {International}
  {Conference} on {World} {Wide} {Web}}} \emph{(\bibinfo{series}{{WWW} '15
  {Companion}})}. \bibinfo{publisher}{Association for Computing Machinery},
  \bibinfo{address}{New York, NY, USA}, \bibinfo{pages}{1185--1188}.
\newblock
\showISBNx{978-1-4503-3473-0}
\urldef\tempurl%
\url{https://doi.org/10.1145/2740908.2741727}
\showDOI{\tempurl}


\bibitem[\protect\citeauthoryear{Campo~Woytuk, Søndergaard, Ciolfi~Felice, and
  Balaam}{Campo~Woytuk et~al\mbox{.}}{2020}]%
        {campo_woytuk_touching_2020}
\bibfield{author}{\bibinfo{person}{Nadia Campo~Woytuk}, \bibinfo{person}{Marie
  Louise~Juul Søndergaard}, \bibinfo{person}{Marianela Ciolfi~Felice}, {and}
  \bibinfo{person}{Madeline Balaam}.} \bibinfo{year}{2020}\natexlab{}.
\newblock \showarticletitle{Touching and {Being} in {Touch} with the
  {Menstruating} {Body}}. In \bibinfo{booktitle}{\emph{Proceedings of the 2020
  {CHI} {Conference} on {Human} {Factors} in {Computing} {Systems}}}
  \emph{(\bibinfo{series}{{CHI} '20})}. \bibinfo{publisher}{Association for
  Computing Machinery}, \bibinfo{address}{New York, NY, USA},
  \bibinfo{pages}{1--14}.
\newblock
\showISBNx{978-1-4503-6708-0}
\urldef\tempurl%
\url{https://doi.org/10.1145/3313831.3376471}
\showDOI{\tempurl}


\bibitem[\protect\citeauthoryear{Chen, Ren, and Riedl}{Chen
  et~al\mbox{.}}{2010}]%
        {chen_effects_2010}
\bibfield{author}{\bibinfo{person}{Jilin Chen}, \bibinfo{person}{Yuqing Ren},
  {and} \bibinfo{person}{John Riedl}.} \bibinfo{year}{2010}\natexlab{}.
\newblock \showarticletitle{The effects of diversity on group productivity and
  member withdrawal in online volunteer groups}. In
  \bibinfo{booktitle}{\emph{Proceedings of the 28th international conference on
  {Human} factors in computing systems}} \emph{(\bibinfo{series}{{CHI} '10})}.
  \bibinfo{publisher}{ACM}, \bibinfo{address}{New York, NY, USA},
  \bibinfo{pages}{821--830}.
\newblock
\showISBNx{978-1-60558-929-9}
\urldef\tempurl%
\url{https://doi.org/10.1145/1753326.1753447}
\showDOI{\tempurl}


\bibitem[\protect\citeauthoryear{Cole}{Cole}{1998}]%
        {cole_poison_1998}
\bibfield{author}{\bibinfo{person}{Leonard~A. Cole}.}
  \bibinfo{year}{1998}\natexlab{}.
\newblock \showarticletitle{The {Poison} {Weapons} {Taboo}: {Biology},
  {Culture}, and {Policy}}.
\newblock \bibinfo{journal}{\emph{Politics and the Life Sciences}}
  \bibinfo{volume}{17}, \bibinfo{number}{2} (\bibinfo{date}{Sept.}
  \bibinfo{year}{1998}), \bibinfo{pages}{119--132}.
\newblock
\showISSN{0730-9384, 1471-5457}
\urldef\tempurl%
\url{https://doi.org/10.1017/S0730938400012119}
\showDOI{\tempurl}


\bibitem[\protect\citeauthoryear{Collier and Bear}{Collier and Bear}{2012}]%
        {collier_conflict_2012}
\bibfield{author}{\bibinfo{person}{Benjamin Collier} {and}
  \bibinfo{person}{Julia Bear}.} \bibinfo{year}{2012}\natexlab{}.
\newblock \showarticletitle{Conflict, criticism, or confidence: an empirical
  examination of the gender gap in {Wikipedia} contributions}. In
  \bibinfo{booktitle}{\emph{Proceedings of the {ACM} 2012 conference on
  {Computer} {Supported} {Cooperative} {Work}}} \emph{(\bibinfo{series}{{CSCW}
  '12})}. \bibinfo{publisher}{ACM}, \bibinfo{address}{New York, NY, USA},
  \bibinfo{pages}{383--392}.
\newblock
\showISBNx{978-1-4503-1086-4}
\urldef\tempurl%
\url{https://doi.org/10.1145/2145204.2145265}
\showDOI{\tempurl}


\bibitem[\protect\citeauthoryear{Crespo~Fernández}{Crespo~Fernández}{2011}]%
        {crespo_fernandez_conceptual_2011}
\bibfield{author}{\bibinfo{person}{Eliecer Crespo~Fernández}.}
  \bibinfo{year}{2011}\natexlab{}.
\newblock \showarticletitle{Conceptual metaphors in taboo-induced lexical
  variation}.
\newblock \bibinfo{journal}{\emph{Revista Alicantina de Estudios Ingleses}}
  \bibinfo{number}{24} (\bibinfo{date}{Nov.} \bibinfo{year}{2011}),
  \bibinfo{pages}{53}.
\newblock
\showISSN{2171-861X, 0214-4808}
\urldef\tempurl%
\url{https://doi.org/10.14198/raei.2011.24.03}
\showDOI{\tempurl}


\bibitem[\protect\citeauthoryear{Douglas}{Douglas}{1978}]%
        {douglas_purity_1978}
\bibfield{author}{\bibinfo{person}{Mary Douglas}.}
  \bibinfo{year}{1978}\natexlab{}.
\newblock \bibinfo{booktitle}{\emph{Purity and danger: an analysis of the
  concepts of pollution and taboo} (\bibinfo{edition}{repr} ed.)}.
\newblock \bibinfo{publisher}{Routledge}, \bibinfo{address}{London}.
\newblock
\showISBNx{978-0-7100-1299-9 978-0-7100-8827-7}


\bibitem[\protect\citeauthoryear{Ford, Sen, Musicant, and Miller}{Ford
  et~al\mbox{.}}{2013}]%
        {ford_getting_2013}
\bibfield{author}{\bibinfo{person}{Heather Ford}, \bibinfo{person}{Shilad Sen},
  \bibinfo{person}{David~R. Musicant}, {and} \bibinfo{person}{Nathaniel
  Miller}.} \bibinfo{year}{2013}\natexlab{}.
\newblock \showarticletitle{Getting to the {Source}: {Where} {Does} {Wikipedia}
  {Get} {Its} {Information} from?}. In \bibinfo{booktitle}{\emph{Proceedings of
  the 9th {International} {Symposium} on {Open} {Collaboration}}}
  \emph{(\bibinfo{series}{{WikiSym} '13})}. \bibinfo{publisher}{ACM},
  \bibinfo{address}{New York, NY, USA}, \bibinfo{pages}{9:1--9:10}.
\newblock
\showISBNx{978-1-4503-1852-5}
\urldef\tempurl%
\url{https://doi.org/10.1145/2491055.2491064}
\showDOI{\tempurl}


\bibitem[\protect\citeauthoryear{Ford and Wajcman}{Ford and Wajcman}{2017}]%
        {ford_anyone_2017}
\bibfield{author}{\bibinfo{person}{Heather Ford} {and} \bibinfo{person}{Judy
  Wajcman}.} \bibinfo{year}{2017}\natexlab{}.
\newblock \showarticletitle{‘{Anyone} can edit’, not everyone does:
  {Wikipedia}’s infrastructure and the gender gap}.
\newblock \bibinfo{journal}{\emph{Social Studies of Science}}
  \bibinfo{volume}{47}, \bibinfo{number}{4} (\bibinfo{year}{2017}),
  \bibinfo{pages}{511--527}.
\newblock
\showISSN{0306-3127}
\urldef\tempurl%
\url{https://doi.org/10.1177/0306312717692172}
\showDOI{\tempurl}


\bibitem[\protect\citeauthoryear{Forte, Andalibi, and Greenstadt}{Forte
  et~al\mbox{.}}{2017}]%
        {forte_privacy_2017}
\bibfield{author}{\bibinfo{person}{Andrea Forte}, \bibinfo{person}{Nazanin
  Andalibi}, {and} \bibinfo{person}{Rachel Greenstadt}.}
  \bibinfo{year}{2017}\natexlab{}.
\newblock \showarticletitle{Privacy, anonymity, and perceived risk in open
  collaboration: a study of {Tor} users and {Wikipedians}}. In
  \bibinfo{booktitle}{\emph{Proceedings of the 2017 {ACM} {Conference} on
  {Computer} {Supported} {Cooperative} {Work} and {Social} {Computing}}}
  \emph{(\bibinfo{series}{{CSCW} '17})}. \bibinfo{publisher}{ACM},
  \bibinfo{address}{New York, NY}, \bibinfo{pages}{1800--1811}.
\newblock
\showISBNx{978-1-4503-4335-0}
\urldef\tempurl%
\url{https://doi.org/10.1145/2998181.2998273}
\showDOI{\tempurl}


\bibitem[\protect\citeauthoryear{Geiger and Halfaker}{Geiger and
  Halfaker}{2017}]%
        {geiger_operationalizing_2017}
\bibfield{author}{\bibinfo{person}{R.~Stuart Geiger} {and}
  \bibinfo{person}{Aaron Halfaker}.} \bibinfo{year}{2017}\natexlab{}.
\newblock \showarticletitle{Operationalizing conflict and cooperation between
  automated software agents in wikipedia: a replication and expansion of `even
  good bots fight'}.
\newblock \bibinfo{journal}{\emph{Proceedings of the ACM on Human-Computer
  Interaction}} \bibinfo{volume}{1}, \bibinfo{number}{CSCW}
  (\bibinfo{date}{Dec.} \bibinfo{year}{2017}), \bibinfo{pages}{1--33}.
\newblock
\showISSN{25730142}
\urldef\tempurl%
\url{https://doi.org/10.1145/3134684}
\showDOI{\tempurl}


\bibitem[\protect\citeauthoryear{Goffman}{Goffman}{1963}]%
        {goffman_stigma_1963}
\bibfield{author}{\bibinfo{person}{Erving Goffman}.}
  \bibinfo{year}{1963}\natexlab{}.
\newblock \bibinfo{booktitle}{\emph{Stigma: {Notes} on the management of
  spoiled identity}}.
\newblock \bibinfo{publisher}{Simon and Schuster}.
\newblock
\showISBNx{1-4391-8833-5}


\bibitem[\protect\citeauthoryear{Gorbatai}{Gorbatai}{2014}]%
        {gorbatai_paradox_2014}
\bibfield{author}{\bibinfo{person}{Andreea~D. Gorbatai}.}
  \bibinfo{year}{2014}\natexlab{}.
\newblock \bibinfo{booktitle}{\emph{The {Paradox} of {Novice} {Contributions}
  to {Collective} {Production}: {Evidence} from {Wikipedia}}}.
\newblock \bibinfo{type}{{T}echnical {R}eport}.
\newblock
\urldef\tempurl%
\url{https://papers.ssrn.com/sol3/papers.cfm?abstract_id=1949327}
\showURL{%
\tempurl}


\bibitem[\protect\citeauthoryear{Greenstein and Zhu}{Greenstein and
  Zhu}{2016}]%
        {greenstein_open_2016}
\bibfield{author}{\bibinfo{person}{Shane Greenstein} {and}
  \bibinfo{person}{Feng Zhu}.} \bibinfo{year}{2016}\natexlab{}.
\newblock \showarticletitle{Open {Content}, {Linus}’ {Law}, and {Neutral}
  {Point} of {View}}.
\newblock \bibinfo{journal}{\emph{Information Systems Research}}
  \bibinfo{volume}{27}, \bibinfo{number}{3} (\bibinfo{date}{Sept.}
  \bibinfo{year}{2016}), \bibinfo{pages}{618--635}.
\newblock
\showISSN{1047-7047, 1526-5536}
\urldef\tempurl%
\url{https://doi.org/10.1287/isre.2016.0643}
\showDOI{\tempurl}


\bibitem[\protect\citeauthoryear{Haklay, Basiouka, Antoniou, and Ather}{Haklay
  et~al\mbox{.}}{2010}]%
        {haklay_how_2010}
\bibfield{author}{\bibinfo{person}{Mordechai~(Muki) Haklay},
  \bibinfo{person}{Sofia Basiouka}, \bibinfo{person}{Vyron Antoniou}, {and}
  \bibinfo{person}{Aamer Ather}.} \bibinfo{year}{2010}\natexlab{}.
\newblock \showarticletitle{How {Many} {Volunteers} {Does} it {Take} to {Map}
  an {Area} {Well}? {The} {Validity} of {Linus}’ {Law} to {Volunteered}
  {Geographic} {Information}}.
\newblock \bibinfo{journal}{\emph{The Cartographic Journal}}
  \bibinfo{volume}{47}, \bibinfo{number}{4} (\bibinfo{date}{Nov.}
  \bibinfo{year}{2010}), \bibinfo{pages}{315--322}.
\newblock
\showISSN{0008-7041, 1743-2774}
\urldef\tempurl%
\url{https://doi.org/10.1179/000870410X12911304958827}
\showDOI{\tempurl}


\bibitem[\protect\citeauthoryear{Halfaker}{Halfaker}{2017}]%
        {halfaker_interpolating_2017}
\bibfield{author}{\bibinfo{person}{Aaron Halfaker}.}
  \bibinfo{year}{2017}\natexlab{}.
\newblock \showarticletitle{Interpolating {Quality} {Dynamics} in {Wikipedia}
  and {Demonstrating} the {Keilana} {Effect}}. \bibinfo{publisher}{ACM Press},
  \bibinfo{pages}{1--9}.
\newblock
\showISBNx{978-1-4503-5187-4}
\urldef\tempurl%
\url{https://doi.org/10.1145/3125433.3125475}
\showDOI{\tempurl}


\bibitem[\protect\citeauthoryear{Halfaker, Kittur, and Riedl}{Halfaker
  et~al\mbox{.}}{2011}]%
        {halfaker_dont_2011}
\bibfield{author}{\bibinfo{person}{Aaron Halfaker}, \bibinfo{person}{Aniket
  Kittur}, {and} \bibinfo{person}{John Riedl}.}
  \bibinfo{year}{2011}\natexlab{}.
\newblock \showarticletitle{Don't bite the newbies: {How} reverts affect the
  quantity and quality of {Wikipedia} work}. In
  \bibinfo{booktitle}{\emph{Proceedings of the 7th {International} {Symposium}
  on {Wikis} and {Open} {Collaboration} ({WikiSym} '11)}}.
  \bibinfo{publisher}{ACM}, \bibinfo{address}{New York, NY},
  \bibinfo{pages}{163--172}.
\newblock
\showISBNx{978-1-4503-0909-7}
\urldef\tempurl%
\url{https://doi.org/10.1145/2038558.2038585}
\showDOI{\tempurl}


\bibitem[\protect\citeauthoryear{Hargittai and Jennrich}{Hargittai and
  Jennrich}{2016}]%
        {hargittai_online_2016}
\bibfield{author}{\bibinfo{person}{Eszter Hargittai} {and}
  \bibinfo{person}{Kaitlin Jennrich}.} \bibinfo{year}{2016}\natexlab{}.
\newblock \showarticletitle{The {Online} {Participation} {Divide}}.
\newblock In \bibinfo{booktitle}{\emph{The {Communication} {Crisis} in
  {America}, {And} {How} to {Fix} {It}}},
  \bibfield{editor}{\bibinfo{person}{Mark Lloyd} {and}
  \bibinfo{person}{Lewis~A. Friedland}} (Eds.). \bibinfo{publisher}{Palgrave
  Macmillan US}, \bibinfo{address}{New York}, \bibinfo{pages}{199--213}.
\newblock
\showISBNx{978-1-349-94925-0}
\urldef\tempurl%
\url{https://doi.org/10.1057/978-1-349-94925-0_13}
\showURL{%
\tempurl}


\bibitem[\protect\citeauthoryear{Hargittai and Shaw}{Hargittai and
  Shaw}{2015}]%
        {hargittai_mind_2015}
\bibfield{author}{\bibinfo{person}{Eszter Hargittai} {and}
  \bibinfo{person}{Aaron Shaw}.} \bibinfo{year}{2015}\natexlab{}.
\newblock \showarticletitle{Mind the skills gap: the role of {Internet}
  know-how and gender in differentiated contributions to {Wikipedia}}.
\newblock \bibinfo{journal}{\emph{Information, Communication \& Society}}
  \bibinfo{volume}{18}, \bibinfo{number}{4} (\bibinfo{date}{April}
  \bibinfo{year}{2015}), \bibinfo{pages}{424--442}.
\newblock
\showISSN{1369-118X}
\urldef\tempurl%
\url{https://doi.org/10.1080/1369118X.2014.957711}
\showDOI{\tempurl}


\bibitem[\protect\citeauthoryear{Hecht and Gergle}{Hecht and Gergle}{2010}]%
        {hecht_tower_2010}
\bibfield{author}{\bibinfo{person}{Brent Hecht} {and} \bibinfo{person}{Darren
  Gergle}.} \bibinfo{year}{2010}\natexlab{}.
\newblock \showarticletitle{The {Tower} of {Babel} meets {Web} 2.0:
  user-generated content and its applications in a multilingual context}. In
  \bibinfo{booktitle}{\emph{Proceedings of the {SIGCHI} conference on human
  factors in computing systems}} \emph{(\bibinfo{series}{{CHI} '10})}.
  \bibinfo{publisher}{ACM}, \bibinfo{address}{Atlanta, Georgia, USA},
  \bibinfo{pages}{291--300}.
\newblock
\showISBNx{978-1-60558-929-9}
\urldef\tempurl%
\url{https://doi.org/10.1145/1753326.1753370}
\showDOI{\tempurl}


\bibitem[\protect\citeauthoryear{Helms}{Helms}{2019}]%
        {helms_you_2019}
\bibfield{author}{\bibinfo{person}{Karey Helms}.}
  \bibinfo{year}{2019}\natexlab{}.
\newblock \showarticletitle{Do {You} {Have} to {Pee}?: {A} {Design} {Space} for
  {Intimate} and {Somatic} {Data}}. In \bibinfo{booktitle}{\emph{Proceedings of
  the 2019 on {Designing} {Interactive} {Systems} {Conference}}}.
  \bibinfo{publisher}{ACM}, \bibinfo{address}{San Diego CA USA},
  \bibinfo{pages}{1209--1222}.
\newblock
\showISBNx{978-1-4503-5850-7}
\urldef\tempurl%
\url{https://doi.org/10.1145/3322276.3322290}
\showDOI{\tempurl}


\bibitem[\protect\citeauthoryear{Hill and Shaw}{Hill and Shaw}{2013}]%
        {hill_wikipedia_2013}
\bibfield{author}{\bibinfo{person}{Benjamin~Mako Hill} {and}
  \bibinfo{person}{Aaron Shaw}.} \bibinfo{year}{2013}\natexlab{}.
\newblock \showarticletitle{The {Wikipedia} gender gap revisited:
  characterizing survey response bias with propensity score estimation}.
\newblock \bibinfo{journal}{\emph{PLoS ONE}} \bibinfo{volume}{8},
  \bibinfo{number}{6} (\bibinfo{date}{June} \bibinfo{year}{2013}),
  \bibinfo{pages}{e65782}.
\newblock
\urldef\tempurl%
\url{https://doi.org/10.1371/journal.pone.0065782}
\showDOI{\tempurl}


\bibitem[\protect\citeauthoryear{Hill and Shaw}{Hill and Shaw}{2014}]%
        {hill_consider_2014}
\bibfield{author}{\bibinfo{person}{Benjamin~Mako Hill} {and}
  \bibinfo{person}{Aaron Shaw}.} \bibinfo{year}{2014}\natexlab{}.
\newblock \showarticletitle{Consider the redirect: a missing dimension of
  {Wikipedia} research}. In \bibinfo{booktitle}{\emph{Proceedings of {The}
  {International} {Symposium} on {Open} {Collaboration}}}
  \emph{(\bibinfo{series}{{OpenSym} '14})}. \bibinfo{publisher}{ACM},
  \bibinfo{address}{New York, NY, USA}, \bibinfo{pages}{28:1--28:4}.
\newblock
\showISBNx{978-1-4503-3016-9}
\urldef\tempurl%
\url{https://doi.org/10.1145/2641580.2641616}
\showDOI{\tempurl}


\bibitem[\protect\citeauthoryear{Hill and Shaw}{Hill and Shaw}{2015}]%
        {hill_page_2015}
\bibfield{author}{\bibinfo{person}{Benjamin~Mako Hill} {and}
  \bibinfo{person}{Aaron Shaw}.} \bibinfo{year}{2015}\natexlab{}.
\newblock \showarticletitle{Page protection: another missing dimension of
  {Wikipedia} research}. In \bibinfo{booktitle}{\emph{Proceedings of the 11th
  {International} {Symposium} on {Open} {Collaboration}}}
  \emph{(\bibinfo{series}{{OpenSym} '15})}. \bibinfo{publisher}{ACM},
  \bibinfo{address}{New York, NY, USA}, \bibinfo{pages}{15:1--15:4}.
\newblock
\showISBNx{978-1-4503-3666-6}
\urldef\tempurl%
\url{https://doi.org/10.1145/2788993.2789846}
\showDOI{\tempurl}


\bibitem[\protect\citeauthoryear{Hill and Shaw}{Hill and Shaw}{2021}]%
        {hill_hidden_2021}
\bibfield{author}{\bibinfo{person}{Benjamin~Mako Hill} {and}
  \bibinfo{person}{Aaron Shaw}.} \bibinfo{year}{2021}\natexlab{}.
\newblock \showarticletitle{The {Hidden} {Costs} of {Requiring} {Accounts}:
  {Quasi}-{Experimental} {Evidence} {From} {Peer} {Production}}.
\newblock \bibinfo{journal}{\emph{Communication Research}}
  \bibinfo{volume}{48}, \bibinfo{number}{6} (\bibinfo{date}{Aug.}
  \bibinfo{year}{2021}), \bibinfo{pages}{771--795}.
\newblock
\showISSN{0093-6502, 1552-3810}
\urldef\tempurl%
\url{https://doi.org/10.1177/0093650220910345}
\showDOI{\tempurl}


\bibitem[\protect\citeauthoryear{Jacobs}{Jacobs}{2019}]%
        {jacobs_wikipedia_2019}
\bibfield{author}{\bibinfo{person}{Julia Jacobs}.}
  \bibinfo{year}{2019}\natexlab{}.
\newblock \showarticletitle{Wikipedia {Isn}’t {Officially} a {Social}
  {Network}. {But} the {Harassment} {Can} {Get} {Ugly}.}
\newblock \bibinfo{journal}{\emph{The New York Times}} (\bibinfo{date}{April}
  \bibinfo{year}{2019}).
\newblock
\showISSN{0362-4331}
\urldef\tempurl%
\url{https://www.nytimes.com/2019/04/08/us/wikipedia-harassment-wikimedia-foundation.html}
\showURL{%
\tempurl}


\bibitem[\protect\citeauthoryear{James}{James}{2016}]%
        {james_wikiproject_2016}
\bibfield{author}{\bibinfo{person}{Richard James}.}
  \bibinfo{year}{2016}\natexlab{}.
\newblock \showarticletitle{{WikiProject} {Medicine}: {Creating} {Credibility}
  in {Consumer} {Health}}.
\newblock \bibinfo{journal}{\emph{Journal of Hospital Librarianship}}
  \bibinfo{volume}{16}, \bibinfo{number}{4} (\bibinfo{date}{Oct.}
  \bibinfo{year}{2016}), \bibinfo{pages}{344--351}.
\newblock
\showISSN{1532-3269, 1532-3277}
\urldef\tempurl%
\url{https://doi.org/10.1080/15323269.2016.1221284}
\showDOI{\tempurl}


\bibitem[\protect\citeauthoryear{Kang, Brown, and Kiesler}{Kang
  et~al\mbox{.}}{2013}]%
        {kang_why_2013}
\bibfield{author}{\bibinfo{person}{Ruogu Kang}, \bibinfo{person}{Stephanie
  Brown}, {and} \bibinfo{person}{Sara Kiesler}.}
  \bibinfo{year}{2013}\natexlab{}.
\newblock \showarticletitle{Why do people seek anonymity on the internet?
  informing policy and design}. In \bibinfo{booktitle}{\emph{Proceedings of the
  {SIGCHI} {Conference} on {Human} {Factors} in {Computing} {Systems}}}
  \emph{(\bibinfo{series}{{CHI} '13})}. \bibinfo{publisher}{Association for
  Computing Machinery}, \bibinfo{address}{New York, NY, USA},
  \bibinfo{pages}{2657--2666}.
\newblock
\showISBNx{978-1-4503-1899-0}
\urldef\tempurl%
\url{https://doi.org/10.1145/2470654.2481368}
\showDOI{\tempurl}


\bibitem[\protect\citeauthoryear{Kannabiran, Bardzell, and Bardzell}{Kannabiran
  et~al\mbox{.}}{2011}]%
        {kannabiran_how_2011}
\bibfield{author}{\bibinfo{person}{Gopinaath Kannabiran},
  \bibinfo{person}{Jeffrey Bardzell}, {and} \bibinfo{person}{Shaowen
  Bardzell}.} \bibinfo{year}{2011}\natexlab{}.
\newblock \showarticletitle{How {HCI} talks about sexuality: discursive
  strategies, blind spots, and opportunities for future research}. In
  \bibinfo{booktitle}{\emph{Proceedings of the 2011 annual conference on
  {Human} factors in computing systems - {CHI} '11}}. \bibinfo{publisher}{ACM
  Press}, \bibinfo{address}{Vancouver, BC, Canada}, \bibinfo{pages}{695}.
\newblock
\showISBNx{978-1-4503-0228-9}
\urldef\tempurl%
\url{https://doi.org/10.1145/1978942.1979043}
\showDOI{\tempurl}


\bibitem[\protect\citeauthoryear{Khatri, Shaw, Dasgupta, and Hill}{Khatri
  et~al\mbox{.}}{2022}]%
        {khatri_social_2022}
\bibfield{author}{\bibinfo{person}{Sejal Khatri}, \bibinfo{person}{Aaron Shaw},
  \bibinfo{person}{Sayamindu Dasgupta}, {and} \bibinfo{person}{Benjamin~Mako
  Hill}.} \bibinfo{year}{2022}\natexlab{}.
\newblock \showarticletitle{The social embeddedness of peer production: {A}
  comparative qualitative analysis of three {Indian} language {Wikipedia}
  editions}. In \bibinfo{booktitle}{\emph{{CHI} {Conference} on {Human}
  {Factors} in {Computing} {Systems}}}. \bibinfo{publisher}{ACM},
  \bibinfo{address}{New Orleans LA USA}, \bibinfo{pages}{1--18}.
\newblock
\showISBNx{978-1-4503-9157-3}
\urldef\tempurl%
\url{https://doi.org/10.1145/3491102.3501832}
\showDOI{\tempurl}


\bibitem[\protect\citeauthoryear{King, Pan, and Roberts}{King
  et~al\mbox{.}}{2013}]%
        {king_how_2013}
\bibfield{author}{\bibinfo{person}{Gary King}, \bibinfo{person}{Jennifer Pan},
  {and} \bibinfo{person}{Margaret~E. Roberts}.}
  \bibinfo{year}{2013}\natexlab{}.
\newblock \showarticletitle{How censorship in {China} allows government
  criticism but silences collective expression}.
\newblock \bibinfo{journal}{\emph{American Political Science Review}}
  \bibinfo{volume}{107}, \bibinfo{number}{2} (\bibinfo{date}{May}
  \bibinfo{year}{2013}).
\newblock
\urldef\tempurl%
\url{https://doi.org/10.1017/S0003055413000014}
\showDOI{\tempurl}


\bibitem[\protect\citeauthoryear{Kraut, Resnick, and Kiesler}{Kraut
  et~al\mbox{.}}{2012}]%
        {kraut_building_2012}
\bibfield{author}{\bibinfo{person}{Robert~E. Kraut}, \bibinfo{person}{Paul
  Resnick}, {and} \bibinfo{person}{Sara Kiesler}.}
  \bibinfo{year}{2012}\natexlab{}.
\newblock \bibinfo{booktitle}{\emph{Building successful online communities:
  {Evidence}-based social design}}.
\newblock \bibinfo{publisher}{MIT Press}, \bibinfo{address}{Cambridge, MA}.
\newblock
\showISBNx{978-0-262-29831-5}


\bibitem[\protect\citeauthoryear{Lazar, Su, Bardzell, and Bardzell}{Lazar
  et~al\mbox{.}}{2019}]%
        {lazar_parting_2019}
\bibfield{author}{\bibinfo{person}{Amanda Lazar},
  \bibinfo{person}{Norman~Makoto Su}, \bibinfo{person}{Jeffrey Bardzell}, {and}
  \bibinfo{person}{Shaowen Bardzell}.} \bibinfo{year}{2019}\natexlab{}.
\newblock \showarticletitle{Parting the {Red} {Sea}: {Sociotechnical} {Systems}
  and {Lived} {Experiences} of {Menopause}}. In
  \bibinfo{booktitle}{\emph{Proceedings of the 2019 {CHI} {Conference} on
  {Human} {Factors} in {Computing} {Systems}}}. \bibinfo{publisher}{ACM},
  \bibinfo{address}{Glasgow Scotland Uk}, \bibinfo{pages}{1--16}.
\newblock
\showISBNx{978-1-4503-5970-2}
\urldef\tempurl%
\url{https://doi.org/10.1145/3290605.3300710}
\showDOI{\tempurl}


\bibitem[\protect\citeauthoryear{Litt}{Litt}{2012}]%
        {litt_knock_2012}
\bibfield{author}{\bibinfo{person}{Eden Litt}.}
  \bibinfo{year}{2012}\natexlab{}.
\newblock \showarticletitle{\textit{{Knock}, {Knock}} . {Who}'s {There}? {The}
  {Imagined} {Audience}}.
\newblock \bibinfo{journal}{\emph{Journal of Broadcasting \& Electronic Media}}
  \bibinfo{volume}{56}, \bibinfo{number}{3} (\bibinfo{date}{July}
  \bibinfo{year}{2012}), \bibinfo{pages}{330--345}.
\newblock
\showISSN{0883-8151, 1550-6878}
\urldef\tempurl%
\url{https://doi.org/10.1080/08838151.2012.705195}
\showDOI{\tempurl}


\bibitem[\protect\citeauthoryear{Malinowski}{Malinowski}{1926}]%
        {malinowski_crime_1926}
\bibfield{author}{\bibinfo{person}{Bronisław Malinowski}.}
  \bibinfo{year}{1926}\natexlab{}.
\newblock \bibinfo{booktitle}{\emph{Crime and custom in savage society}}.
\newblock \bibinfo{publisher}{Harcourt Brace \& Company}, \bibinfo{address}{New
  York}.
\newblock


\bibitem[\protect\citeauthoryear{Marwick and boyd}{Marwick and boyd}{2011}]%
        {marwick_i_2011}
\bibfield{author}{\bibinfo{person}{A.~E. Marwick} {and} \bibinfo{person}{danah
  boyd}.} \bibinfo{year}{2011}\natexlab{}.
\newblock \showarticletitle{I tweet honestly, {I} tweet passionately: {Twitter}
  users, context collapse, and the imagined audience}.
\newblock \bibinfo{journal}{\emph{New Media \& Society}} \bibinfo{volume}{13},
  \bibinfo{number}{1} (\bibinfo{date}{Feb.} \bibinfo{year}{2011}),
  \bibinfo{pages}{114--133}.
\newblock
\showISSN{1461-4448}
\urldef\tempurl%
\url{https://doi.org/10.1177/1461444810365313}
\showDOI{\tempurl}


\bibitem[\protect\citeauthoryear{Marx}{Marx}{2016}]%
        {marx_windows_2016}
\bibfield{author}{\bibinfo{person}{Gary~T. Marx}.}
  \bibinfo{year}{2016}\natexlab{}.
\newblock \bibinfo{booktitle}{\emph{Windows into the soul: surveillance and
  society in an age of high technology}}.
\newblock \bibinfo{publisher}{The University of Chicago Press},
  \bibinfo{address}{Chicago ; London}.
\newblock
\showISBNx{978-0-226-28588-7 978-0-226-28591-7}


\bibitem[\protect\citeauthoryear{McGlone and Batchelor}{McGlone and
  Batchelor}{2003}]%
        {mcglone_looking_2003}
\bibfield{author}{\bibinfo{person}{Matthew~S. McGlone} {and}
  \bibinfo{person}{Jennifer~A. Batchelor}.} \bibinfo{year}{2003}\natexlab{}.
\newblock \showarticletitle{Looking {Out} for {Number} {One}: {Euphemism} and
  {Face}}.
\newblock \bibinfo{journal}{\emph{Journal of Communication}}
  \bibinfo{volume}{53}, \bibinfo{number}{2} (\bibinfo{date}{June}
  \bibinfo{year}{2003}), \bibinfo{pages}{251--264}.
\newblock
\showISSN{0021-9916, 1460-2466}
\urldef\tempurl%
\url{https://doi.org/10.1111/j.1460-2466.2003.tb02589.x}
\showDOI{\tempurl}


\bibitem[\protect\citeauthoryear{Menking, Erickson, and Pratt}{Menking
  et~al\mbox{.}}{2019}]%
        {menking_people_2019}
\bibfield{author}{\bibinfo{person}{Amanda Menking}, \bibinfo{person}{Ingrid
  Erickson}, {and} \bibinfo{person}{Wanda Pratt}.}
  \bibinfo{year}{2019}\natexlab{}.
\newblock \showarticletitle{People who can take it: how women {Wikipedians}
  negotiate and navigate safety}. In \bibinfo{booktitle}{\emph{Proceedings of
  the 2019 {CHI} {Conference} on {Human} {Factors} in {Computing} {Systems}}}
  \emph{(\bibinfo{series}{{CHI} '19})}. \bibinfo{publisher}{Association for
  Computing Machinery}, \bibinfo{address}{Glasgow, Scotland, UK},
  \bibinfo{pages}{472:1--472:14}.
\newblock
\showISBNx{978-1-4503-5970-2}
\urldef\tempurl%
\url{https://doi.org/10.1145/3290605.3300702}
\showDOI{\tempurl}


\bibitem[\protect\citeauthoryear{Miquel-Ribé and Laniado}{Miquel-Ribé and
  Laniado}{2020}]%
        {miquel-ribe_wikipedia_2020}
\bibfield{author}{\bibinfo{person}{Marc Miquel-Ribé} {and}
  \bibinfo{person}{David Laniado}.} \bibinfo{year}{2020}\natexlab{}.
\newblock \showarticletitle{The {Wikipedia} {Diversity} {Observatory}: {A}
  {Project} to {Identify} and {Bridge} {Content} {Gaps} in {Wikipedia}}. In
  \bibinfo{booktitle}{\emph{Proceedings of the 16th {International} {Symposium}
  on {Open} {Collaboration}}}. \bibinfo{publisher}{ACM},
  \bibinfo{address}{Virtual conference Spain}, \bibinfo{pages}{1--4}.
\newblock
\showISBNx{978-1-4503-8779-8}
\urldef\tempurl%
\url{https://doi.org/10.1145/3412569.3412866}
\showDOI{\tempurl}


\bibitem[\protect\citeauthoryear{Moitra, Hassan, Mandal, Bhuiyan, and
  Ahmed}{Moitra et~al\mbox{.}}{2020}]%
        {moitra_understanding_2020}
\bibfield{author}{\bibinfo{person}{Aparna Moitra}, \bibinfo{person}{Naeemul
  Hassan}, \bibinfo{person}{Manash~Kumar Mandal}, \bibinfo{person}{Mansurul
  Bhuiyan}, {and} \bibinfo{person}{Syed~Ishtiaque Ahmed}.}
  \bibinfo{year}{2020}\natexlab{}.
\newblock \showarticletitle{Understanding the {Challenges} for {Bangladeshi}
  {Women} to {Participate} in \#{MeToo} {Movement}}.
\newblock \bibinfo{journal}{\emph{Proceedings of the ACM on Human-Computer
  Interaction}} \bibinfo{volume}{4}, \bibinfo{number}{GROUP}
  (\bibinfo{date}{Jan.} \bibinfo{year}{2020}), \bibinfo{pages}{1--25}.
\newblock
\showISSN{2573-0142}
\urldef\tempurl%
\url{https://doi.org/10.1145/3375195}
\showDOI{\tempurl}


\bibitem[\protect\citeauthoryear{Morgan, Gilbert, McDonald, and Zachry}{Morgan
  et~al\mbox{.}}{2013}]%
        {morgan_project_2013}
\bibfield{author}{\bibinfo{person}{Jonathan~T. Morgan},
  \bibinfo{person}{Michael Gilbert}, \bibinfo{person}{David~W. McDonald}, {and}
  \bibinfo{person}{Mark Zachry}.} \bibinfo{year}{2013}\natexlab{}.
\newblock \showarticletitle{Project talk: coordination work and group
  membership in {WikiProjects}}. In \bibinfo{booktitle}{\emph{Proceedings of
  the 9th {International} {Symposium} on {Open} {Collaboration}}}.
  \bibinfo{publisher}{ACM}, \bibinfo{address}{Hong Kong China},
  \bibinfo{pages}{1--10}.
\newblock
\showISBNx{978-1-4503-1852-5}
\urldef\tempurl%
\url{https://doi.org/10.1145/2491055.2491058}
\showDOI{\tempurl}


\bibitem[\protect\citeauthoryear{Nafus}{Nafus}{2012}]%
        {nafus_patches_2012}
\bibfield{author}{\bibinfo{person}{Dawn Nafus}.}
  \bibinfo{year}{2012}\natexlab{}.
\newblock \showarticletitle{‘{Patches} don’t have gender’: {What} is not
  open in open source software}.
\newblock \bibinfo{journal}{\emph{New Media \& Society}} \bibinfo{volume}{14},
  \bibinfo{number}{4} (\bibinfo{date}{June} \bibinfo{year}{2012}),
  \bibinfo{pages}{669--683}.
\newblock
\showISSN{1461-4448}
\urldef\tempurl%
\url{https://doi.org/10.1177/1461444811422887}
\showURL{%
\tempurl}


\bibitem[\protect\citeauthoryear{Paling}{Paling}{2015}]%
        {paling_wikipedias_2015}
\bibfield{author}{\bibinfo{person}{Emma Paling}.}
  \bibinfo{year}{2015}\natexlab{}.
\newblock \bibinfo{title}{Wikipedia's {Hostility} to {Women}}.
\newblock
\newblock
\urldef\tempurl%
\url{https://www.theatlantic.com/technology/archive/2015/10/how-wikipedia-is-hostile-to-women/411619/}
\showURL{%
\tempurl}
\newblock
\shownote{Section: Technology.}


\bibitem[\protect\citeauthoryear{Pedregosa, Varoquaux, Gramfort, Michel,
  Thirion, Grisel, Blondel, Prettenhofer, Weiss, Dubourg, Vanderplas, Passos,
  Cournapeau, Brucher, Perrot, and Duchesnay}{Pedregosa et~al\mbox{.}}{2011}]%
        {pedregosa_scikit-learn:_2011}
\bibfield{author}{\bibinfo{person}{Fabian Pedregosa}, \bibinfo{person}{Gaël
  Varoquaux}, \bibinfo{person}{Alexandre Gramfort}, \bibinfo{person}{Vincent
  Michel}, \bibinfo{person}{Bertrand Thirion}, \bibinfo{person}{Olivier
  Grisel}, \bibinfo{person}{Mathieu Blondel}, \bibinfo{person}{Peter
  Prettenhofer}, \bibinfo{person}{Ron Weiss}, \bibinfo{person}{Vincent
  Dubourg}, \bibinfo{person}{Jake Vanderplas}, \bibinfo{person}{Alexandre
  Passos}, \bibinfo{person}{David Cournapeau}, \bibinfo{person}{Matthieu
  Brucher}, \bibinfo{person}{Matthieu Perrot}, {and} \bibinfo{person}{Édouard
  Duchesnay}.} \bibinfo{year}{2011}\natexlab{}.
\newblock \showarticletitle{Scikit-learn: {Machine} learning in python}.
\newblock \bibinfo{journal}{\emph{Journal of Machine Learning Research}}
  \bibinfo{volume}{12}, \bibinfo{number}{85} (\bibinfo{date}{Oct.}
  \bibinfo{year}{2011}), \bibinfo{pages}{2825--2830}.
\newblock
\urldef\tempurl%
\url{http://jmlr.csail.mit.edu/papers/v12/pedregosa11a.html}
\showURL{%
\tempurl}


\bibitem[\protect\citeauthoryear{Price}{Price}{2007}]%
        {price_chemical_2007}
\bibfield{author}{\bibinfo{person}{Richard~M. Price}.}
  \bibinfo{year}{2007}\natexlab{}.
\newblock \bibinfo{booktitle}{\emph{The {Chemical} {Weapons} {Taboo}}}.
\newblock \bibinfo{publisher}{Cornell University Press}.
\newblock
\showISBNx{978-1-5017-2954-6}
\urldef\tempurl%
\url{https://www.degruyter.com/document/doi/10.7591/9781501729546/html}
\showURL{%
\tempurl}


\bibitem[\protect\citeauthoryear{Rahman, Rahman, Tripto, Ali, Apon, and
  Shahriyar}{Rahman et~al\mbox{.}}{2021}]%
        {rahman_adolescentbot_2021}
\bibfield{author}{\bibinfo{person}{Rifat Rahman}, \bibinfo{person}{Md.~Rishadur
  Rahman}, \bibinfo{person}{Nafis~Irtiza Tripto},
  \bibinfo{person}{Mohammed~Eunus Ali}, \bibinfo{person}{Sajid~Hasan Apon},
  {and} \bibinfo{person}{Rifat Shahriyar}.} \bibinfo{year}{2021}\natexlab{}.
\newblock \showarticletitle{Adolescentbot: understanding opportunities for
  chatbots in combating adolescent sexual and reproductive health problems in
  bangladesh}. In \bibinfo{booktitle}{\emph{Proceedings of the 2021 {CHI}
  {Conference} on {Human} {Factors} in {Computing} {Systems}}}.
  \bibinfo{publisher}{ACM}, \bibinfo{address}{Yokohama Japan},
  \bibinfo{pages}{1--15}.
\newblock
\showISBNx{978-1-4503-8096-6}
\urldef\tempurl%
\url{https://doi.org/10.1145/3411764.3445694}
\showDOI{\tempurl}


\bibitem[\protect\citeauthoryear{Rajaraman and Ullman}{Rajaraman and
  Ullman}{2011}]%
        {rajaraman_mining_2011}
\bibfield{author}{\bibinfo{person}{Anand Rajaraman} {and}
  \bibinfo{person}{Jeffrey~David Ullman}.} \bibinfo{year}{2011}\natexlab{}.
\newblock \bibinfo{booktitle}{\emph{Mining of {Massive} {Datasets}}}.
\newblock \bibinfo{publisher}{Cambridge University Press},
  \bibinfo{address}{Cambridge}.
\newblock
\showISBNx{978-1-139-05845-2}
\urldef\tempurl%
\url{https://doi.org/10.1017/CBO9781139058452}
\showDOI{\tempurl}


\bibitem[\protect\citeauthoryear{Raymond}{Raymond}{2001}]%
        {raymond_cathedral_2001}
\bibfield{author}{\bibinfo{person}{Eric Raymond}.}
  \bibinfo{year}{2001}\natexlab{}.
\newblock \bibinfo{booktitle}{\emph{The {Cathedral} and the {Bazaar}: {Musings}
  on {Linux} and {Open} {Source} by an {Accidental} {Revolutionary}}}.
\newblock \bibinfo{publisher}{O'Reilly \& Associates},
  \bibinfo{address}{Sebastopol, CA}.
\newblock


\bibitem[\protect\citeauthoryear{Reagle}{Reagle}{2013}]%
        {reagle_free_2013}
\bibfield{author}{\bibinfo{person}{Joseph Reagle}.}
  \bibinfo{year}{2013}\natexlab{}.
\newblock \showarticletitle{“{Free} as in sexist?” {Free} culture and the
  gender gap}.
\newblock \bibinfo{journal}{\emph{First Monday}} \bibinfo{volume}{18},
  \bibinfo{number}{1} (\bibinfo{year}{2013}).
\newblock
\showISSN{13960466}
\urldef\tempurl%
\url{http://firstmonday.org/ojs/index.php/fm/article/view/4291}
\showURL{%
\tempurl}


\bibitem[\protect\citeauthoryear{Rizoiu, Xie, Caetano, and Cebrian}{Rizoiu
  et~al\mbox{.}}{2016}]%
        {rizoiu_evolution_2016}
\bibfield{author}{\bibinfo{person}{Marian-Andrei Rizoiu},
  \bibinfo{person}{Lexing Xie}, \bibinfo{person}{Tiberio Caetano}, {and}
  \bibinfo{person}{Manuel Cebrian}.} \bibinfo{year}{2016}\natexlab{}.
\newblock \showarticletitle{Evolution of {Privacy} {Loss} in {Wikipedia}}. In
  \bibinfo{booktitle}{\emph{Proceedings of the {Ninth} {ACM} {International}
  {Conference} on {Web} {Search} and {Data} {Mining}}}.
  \bibinfo{publisher}{ACM}, \bibinfo{address}{San Francisco California USA},
  \bibinfo{pages}{215--224}.
\newblock
\showISBNx{978-1-4503-3716-8}
\urldef\tempurl%
\url{https://doi.org/10.1145/2835776.2835798}
\showDOI{\tempurl}


\bibitem[\protect\citeauthoryear{Sannon, Bazarova, and Cosley}{Sannon
  et~al\mbox{.}}{2018}]%
        {sannon_privacy_2018}
\bibfield{author}{\bibinfo{person}{Shruti Sannon}, \bibinfo{person}{Natalya~N.
  Bazarova}, {and} \bibinfo{person}{Dan Cosley}.}
  \bibinfo{year}{2018}\natexlab{}.
\newblock \showarticletitle{Privacy {Lies}: {Understanding} {How}, {When}, and
  {Why} {People} {Lie} to {Protect} {Their} {Privacy} in {Multiple} {Online}
  {Contexts}}. In \bibinfo{booktitle}{\emph{Proceedings of the 2018 {CHI}
  {Conference} on {Human} {Factors} in {Computing} {Systems}}}.
  \bibinfo{publisher}{ACM}, \bibinfo{address}{Montreal QC Canada},
  \bibinfo{pages}{1--13}.
\newblock
\showISBNx{978-1-4503-5620-6}
\urldef\tempurl%
\url{https://doi.org/10.1145/3173574.3173626}
\showDOI{\tempurl}


\bibitem[\protect\citeauthoryear{Schelling}{Schelling}{2007}]%
        {schelling_nuclear_2007}
\bibfield{author}{\bibinfo{person}{Thomas~C. Schelling}.}
  \bibinfo{year}{2007}\natexlab{}.
\newblock \bibinfo{title}{The {Nuclear} {Taboo} - {Spring} 2007 - {MIT}
  {International} {Review}}.
\newblock
\newblock
\urldef\tempurl%
\url{http://web.mit.edu/mitir/2007/spring/taboo.html}
\showURL{%
\tempurl}


\bibitem[\protect\citeauthoryear{Semaan, Dosono, and Britton}{Semaan
  et~al\mbox{.}}{2017}]%
        {semaan_impression_2017}
\bibfield{author}{\bibinfo{person}{Bryan Semaan}, \bibinfo{person}{Bryan
  Dosono}, {and} \bibinfo{person}{Lauren~M. Britton}.}
  \bibinfo{year}{2017}\natexlab{}.
\newblock \showarticletitle{Impression {Management} in {High} {Context}
  {Societies}: '{Saving} {Face}' with {ICT}}. In
  \bibinfo{booktitle}{\emph{Proceedings of the 2017 {ACM} {Conference} on
  {Computer} {Supported} {Cooperative} {Work} and {Social} {Computing}}}.
  \bibinfo{publisher}{ACM}, \bibinfo{address}{Portland Oregon USA},
  \bibinfo{pages}{712--725}.
\newblock
\showISBNx{978-1-4503-4335-0}
\urldef\tempurl%
\url{https://doi.org/10.1145/2998181.2998222}
\showDOI{\tempurl}


\bibitem[\protect\citeauthoryear{Shaw and Hargittai}{Shaw and
  Hargittai}{2018}]%
        {shaw_pipeline_2018}
\bibfield{author}{\bibinfo{person}{Aaron Shaw} {and} \bibinfo{person}{Eszter
  Hargittai}.} \bibinfo{year}{2018}\natexlab{}.
\newblock \showarticletitle{The {Pipeline} of {Online} {Participation}
  {Inequalities}: {The} {Case} of {Wikipedia} {Editing}}.
\newblock \bibinfo{journal}{\emph{Journal of Communication}}
  \bibinfo{volume}{68}, \bibinfo{number}{1} (\bibinfo{date}{Feb.}
  \bibinfo{year}{2018}), \bibinfo{pages}{143--168}.
\newblock
\showISSN{0021-9916}
\urldef\tempurl%
\url{https://doi.org/10.1093/joc/jqx003}
\showDOI{\tempurl}


\bibitem[\protect\citeauthoryear{Song}{Song}{2022}]%
        {song_top_2022}
\bibfield{author}{\bibinfo{person}{Victoria Song}.}
  \bibinfo{year}{2022}\natexlab{}.
\newblock \bibinfo{title}{A top {Wikipedia} editor has been arrested in
  {Belarus}}.
\newblock
\newblock
\urldef\tempurl%
\url{https://www.theverge.com/2022/3/11/22973293/wikipedia-editor-russia-belarus-ukraine}
\showURL{%
\tempurl}


\bibitem[\protect\citeauthoryear{Sorcar, Strauber, Loyalka, Kumar, and
  Goldman}{Sorcar et~al\mbox{.}}{2017}]%
        {sorcar_sidestepping_2017}
\bibfield{author}{\bibinfo{person}{Piya Sorcar}, \bibinfo{person}{Benjamin
  Strauber}, \bibinfo{person}{Prashant Loyalka}, \bibinfo{person}{Neha Kumar},
  {and} \bibinfo{person}{Shelley Goldman}.} \bibinfo{year}{2017}\natexlab{}.
\newblock \showarticletitle{Sidestepping the {Elephant} in the {Classroom}:
  {Using} {Culturally} {Localized} {Technology} {To} {Teach} {Around}
  {Taboos}}. In \bibinfo{booktitle}{\emph{Proceedings of the 2017 {CHI}
  {Conference} on {Human} {Factors} in {Computing} {Systems}}}.
  \bibinfo{publisher}{ACM}, \bibinfo{address}{Denver Colorado USA},
  \bibinfo{pages}{2792--2804}.
\newblock
\showISBNx{978-1-4503-4655-9}
\urldef\tempurl%
\url{https://doi.org/10.1145/3025453.3025958}
\showDOI{\tempurl}


\bibitem[\protect\citeauthoryear{Suler}{Suler}{2004}]%
        {suler_online_2004}
\bibfield{author}{\bibinfo{person}{John Suler}.}
  \bibinfo{year}{2004}\natexlab{}.
\newblock \showarticletitle{The online disinhibition effect}.
\newblock \bibinfo{journal}{\emph{CyberPsychology \& Behavior}}
  \bibinfo{volume}{7}, \bibinfo{number}{3} (\bibinfo{date}{June}
  \bibinfo{year}{2004}), \bibinfo{pages}{321--326}.
\newblock
\showISSN{1094-9313, 1557-8364}
\urldef\tempurl%
\url{https://doi.org/10.1089/1094931041291295}
\showDOI{\tempurl}


\bibitem[\protect\citeauthoryear{Sundara~Raman, Shenoy, Kohls, and
  Ensafi}{Sundara~Raman et~al\mbox{.}}{2020}]%
        {sundara_raman_censored_2020}
\bibfield{author}{\bibinfo{person}{Ram Sundara~Raman}, \bibinfo{person}{Prerana
  Shenoy}, \bibinfo{person}{Katharina Kohls}, {and} \bibinfo{person}{Roya
  Ensafi}.} \bibinfo{year}{2020}\natexlab{}.
\newblock \showarticletitle{Censored {Planet}: {An} {Internet}-wide,
  {Longitudinal} {Censorship} {Observatory}}. In
  \bibinfo{booktitle}{\emph{Proceedings of the 2020 {ACM} {SIGSAC} {Conference}
  on {Computer} and {Communications} {Security}}}. \bibinfo{publisher}{ACM},
  \bibinfo{address}{Virtual Event USA}, \bibinfo{pages}{49--66}.
\newblock
\showISBNx{978-1-4503-7089-9}
\urldef\tempurl%
\url{https://doi.org/10.1145/3372297.3417883}
\showDOI{\tempurl}


\bibitem[\protect\citeauthoryear{Søndergaard, Ciolfi~Felice, and
  Balaam}{Søndergaard et~al\mbox{.}}{2021}]%
        {sondergaard_designing_2021}
\bibfield{author}{\bibinfo{person}{Marie Louise~Juul Søndergaard},
  \bibinfo{person}{Marianela Ciolfi~Felice}, {and} \bibinfo{person}{Madeline
  Balaam}.} \bibinfo{year}{2021}\natexlab{}.
\newblock \showarticletitle{Designing {Menstrual} {Technologies} with
  {Adolescents}}. In \bibinfo{booktitle}{\emph{Proceedings of the 2021 {CHI}
  {Conference} on {Human} {Factors} in {Computing} {Systems}}}.
  \bibinfo{publisher}{ACM}, \bibinfo{address}{Yokohama Japan},
  \bibinfo{pages}{1--14}.
\newblock
\showISBNx{978-1-4503-8096-6}
\urldef\tempurl%
\url{https://doi.org/10.1145/3411764.3445471}
\showDOI{\tempurl}


\bibitem[\protect\citeauthoryear{Taber and Whittaker}{Taber and
  Whittaker}{2020}]%
        {taber_finsta_2020}
\bibfield{author}{\bibinfo{person}{Lee Taber} {and} \bibinfo{person}{Steve
  Whittaker}.} \bibinfo{year}{2020}\natexlab{}.
\newblock \showarticletitle{"{On} {Finsta}, {I} can say '{Hail} {Satan}'":
  {Being} {Authentic} but {Disagreeable} on {Instagram}}. In
  \bibinfo{booktitle}{\emph{Proceedings of the 2020 {CHI} {Conference} on
  {Human} {Factors} in {Computing} {Systems}}}. \bibinfo{publisher}{ACM},
  \bibinfo{address}{Honolulu HI USA}, \bibinfo{pages}{1--14}.
\newblock
\showISBNx{978-1-4503-6708-0}
\urldef\tempurl%
\url{https://doi.org/10.1145/3313831.3376182}
\showDOI{\tempurl}


\bibitem[\protect\citeauthoryear{TeBlunthuis}{TeBlunthuis}{2021}]%
        {teblunthuis_measuring_2021}
\bibfield{author}{\bibinfo{person}{Nathan TeBlunthuis}.}
  \bibinfo{year}{2021}\natexlab{}.
\newblock \showarticletitle{Measuring {Wikipedia} {Article} {Quality} in {One}
  {Dimension} by {Extending} {ORES} with {Ordinal} {Regression}}. In
  \bibinfo{booktitle}{\emph{17th {International} {Symposium} on {Open}
  {Collaboration}}}. \bibinfo{publisher}{ACM}, \bibinfo{address}{Online,
  Spain}.
\newblock
\showISBNx{978-1-4503-8500-8}
\urldef\tempurl%
\url{https://doi.org/10.1145/3479986.3479991}
\showDOI{\tempurl}


\bibitem[\protect\citeauthoryear{TeBlunthuis, Hill, and Halfaker}{TeBlunthuis
  et~al\mbox{.}}{2021}]%
        {teblunthuis_effects_2021}
\bibfield{author}{\bibinfo{person}{Nathan TeBlunthuis},
  \bibinfo{person}{Benjamin~Mako Hill}, {and} \bibinfo{person}{Aaron
  Halfaker}.} \bibinfo{year}{2021}\natexlab{}.
\newblock \showarticletitle{Effects of {Algorithmic} {Flagging} on {Fairness}:
  {Quasi}-experimental {Evidence} from {Wikipedia}}.
\newblock \bibinfo{journal}{\emph{Proceedings of the ACM on Human-Computer
  Interaction}} \bibinfo{volume}{5}, \bibinfo{number}{CSCW1}
  (\bibinfo{date}{April} \bibinfo{year}{2021}), \bibinfo{pages}{56:1--56:27}.
\newblock
\urldef\tempurl%
\url{https://doi.org/10.1145/3449130}
\showDOI{\tempurl}


\bibitem[\protect\citeauthoryear{Thebault-Spieker, Hecht, and
  Terveen}{Thebault-Spieker et~al\mbox{.}}{2018}]%
        {thebault-spieker_geographic_2018}
\bibfield{author}{\bibinfo{person}{Jacob Thebault-Spieker},
  \bibinfo{person}{Brent Hecht}, {and} \bibinfo{person}{Loren Terveen}.}
  \bibinfo{year}{2018}\natexlab{}.
\newblock \showarticletitle{Geographic {Biases} are `{Born}, not {Made}':
  {Exploring} {Contributors}' {Spatiotemporal} {Behavior} in {OpenStreetMap}}.
  In \bibinfo{booktitle}{\emph{Proceedings of the 2018 {ACM} {Conference} on
  {Supporting} {Groupwork}}}. \bibinfo{publisher}{ACM Press},
  \bibinfo{pages}{71--82}.
\newblock
\showISBNx{978-1-4503-5562-9}
\urldef\tempurl%
\url{https://doi.org/10.1145/3148330.3148350}
\showDOI{\tempurl}


\bibitem[\protect\citeauthoryear{Thornton and McDonald}{Thornton and
  McDonald}{2012}]%
        {thornton_tagging_2012}
\bibfield{author}{\bibinfo{person}{Katherine Thornton} {and}
  \bibinfo{person}{David~W. McDonald}.} \bibinfo{year}{2012}\natexlab{}.
\newblock \showarticletitle{Tagging {Wikipedia}: collaboratively creating a
  category system}. In \bibinfo{booktitle}{\emph{Proceedings of the 17th {ACM}
  international conference on {Supporting} group work - {GROUP} '12}}.
  \bibinfo{publisher}{ACM Press}, \bibinfo{address}{Sanibel Island, Florida,
  USA}, \bibinfo{pages}{219}.
\newblock
\showISBNx{978-1-4503-1486-2}
\urldef\tempurl%
\url{https://doi.org/10.1145/2389176.2389210}
\showDOI{\tempurl}


\bibitem[\protect\citeauthoryear{Toch, Wang, and Cranor}{Toch
  et~al\mbox{.}}{2012}]%
        {toch_personalization_2012}
\bibfield{author}{\bibinfo{person}{Eran Toch}, \bibinfo{person}{Yang Wang},
  {and} \bibinfo{person}{Lorrie~Faith Cranor}.}
  \bibinfo{year}{2012}\natexlab{}.
\newblock \showarticletitle{Personalization and privacy: a survey of privacy
  risks and remedies in personalization-based systems}.
\newblock \bibinfo{journal}{\emph{User Modeling and User-Adapted Interaction}}
  \bibinfo{volume}{22}, \bibinfo{number}{1-2} (\bibinfo{date}{April}
  \bibinfo{year}{2012}), \bibinfo{pages}{203--220}.
\newblock
\showISSN{0924-1868, 1573-1391}
\urldef\tempurl%
\url{https://doi.org/10.1007/s11257-011-9110-z}
\showDOI{\tempurl}


\bibitem[\protect\citeauthoryear{Tran, Champion, Forte, Hill, and
  Greenstadt}{Tran et~al\mbox{.}}{2020}]%
        {tran_are_2020}
\bibfield{author}{\bibinfo{person}{Chau Tran}, \bibinfo{person}{Kaylea
  Champion}, \bibinfo{person}{Andrea Forte}, \bibinfo{person}{Benjamin~Mako
  Hill}, {and} \bibinfo{person}{Rachel Greenstadt}.}
  \bibinfo{year}{2020}\natexlab{}.
\newblock \showarticletitle{Are anonymity-seekers just like everybody else?
  {An} analysis of contributions to {Wikipedia} from {Tor}}. In
  \bibinfo{booktitle}{\emph{2020 {IEEE} {Symposium} on {Security} and {Privacy}
  ({SP})}}, Vol.~\bibinfo{volume}{1}. \bibinfo{publisher}{IEEE Computer
  Society}, \bibinfo{address}{San Francisco, California},
  \bibinfo{pages}{974--990}.
\newblock
\urldef\tempurl%
\url{https://doi.org/10.1109/SP40000.2020.00053}
\showDOI{\tempurl}


\bibitem[\protect\citeauthoryear{Trevena}{Trevena}{2011}]%
        {trevena_wikiproject_2011}
\bibfield{author}{\bibinfo{person}{L. Trevena}.}
  \bibinfo{year}{2011}\natexlab{}.
\newblock \showarticletitle{{WikiProject} {Medicine}}.
\newblock \bibinfo{journal}{\emph{BMJ}} \bibinfo{volume}{342},
  \bibinfo{number}{jun08 3} (\bibinfo{date}{June} \bibinfo{year}{2011}),
  \bibinfo{pages}{d3387--d3387}.
\newblock
\showISSN{0959-8138, 1468-5833}
\urldef\tempurl%
\url{https://doi.org/10.1136/bmj.d3387}
\showDOI{\tempurl}


\bibitem[\protect\citeauthoryear{Tripodi}{Tripodi}{2021}]%
        {tripodi_ms_2021}
\bibfield{author}{\bibinfo{person}{Francesca Tripodi}.}
  \bibinfo{year}{2021}\natexlab{}.
\newblock \showarticletitle{Ms. {Categorized}: {Gender}, notability, and
  inequality on {Wikipedia}}.
\newblock \bibinfo{journal}{\emph{New Media \& Society}} (\bibinfo{date}{June}
  \bibinfo{year}{2021}), \bibinfo{pages}{146144482110237}.
\newblock
\showISSN{1461-4448, 1461-7315}
\urldef\tempurl%
\url{https://doi.org/10.1177/14614448211023772}
\showDOI{\tempurl}


\bibitem[\protect\citeauthoryear{Tuli, Chopra, Kumar, and Singh}{Tuli
  et~al\mbox{.}}{2018}]%
        {tuli_learning_2018}
\bibfield{author}{\bibinfo{person}{Anupriya Tuli}, \bibinfo{person}{Shaan
  Chopra}, \bibinfo{person}{Neha Kumar}, {and} \bibinfo{person}{Pushpendra
  Singh}.} \bibinfo{year}{2018}\natexlab{}.
\newblock \showarticletitle{Learning \textit{from} and \textit{with}
  {Menstrupedia}: {Towards} {Menstrual} {Health} {Education} in {India}}.
\newblock \bibinfo{journal}{\emph{Proceedings of the ACM on Human-Computer
  Interaction}} \bibinfo{volume}{2}, \bibinfo{number}{CSCW}
  (\bibinfo{date}{Nov.} \bibinfo{year}{2018}), \bibinfo{pages}{1--20}.
\newblock
\showISSN{2573-0142}
\urldef\tempurl%
\url{https://doi.org/10.1145/3274443}
\showDOI{\tempurl}


\bibitem[\protect\citeauthoryear{Vincent, Johnson, and Hecht}{Vincent
  et~al\mbox{.}}{2018}]%
        {vincent_examining_2018}
\bibfield{author}{\bibinfo{person}{Nicholas Vincent}, \bibinfo{person}{Isaac
  Johnson}, {and} \bibinfo{person}{Brent Hecht}.}
  \bibinfo{year}{2018}\natexlab{}.
\newblock \showarticletitle{Examining {Wikipedia} {With} a {Broader} {Lens}:
  {Quantifying} the {Value} of {Wikipedia}'s {Relationships} with {Other}
  {Large}-{Scale} {Online} {Communities}}. \bibinfo{publisher}{ACM Press},
  \bibinfo{pages}{1--13}.
\newblock
\showISBNx{978-1-4503-5620-6}
\urldef\tempurl%
\url{https://doi.org/10.1145/3173574.3174140}
\showDOI{\tempurl}


\bibitem[\protect\citeauthoryear{Warncke-Wang, Ranjan, Terveen, and
  Hecht}{Warncke-Wang et~al\mbox{.}}{2015}]%
        {warncke-wang_misalignment_2015}
\bibfield{author}{\bibinfo{person}{Morten Warncke-Wang}, \bibinfo{person}{Vivek
  Ranjan}, \bibinfo{person}{Loren Terveen}, {and} \bibinfo{person}{Brent
  Hecht}.} \bibinfo{year}{2015}\natexlab{}.
\newblock \showarticletitle{Misalignment between supply and demand of quality
  content in peer production communities}. In
  \bibinfo{booktitle}{\emph{Proceedings of the {Ninth} {International} {AAAI}
  {Conference} on {Web} and {Social} {Media} ({ICWSM} '15)}}.
  \bibinfo{pages}{493--502}.
\newblock


\bibitem[\protect\citeauthoryear{Yang, Halfaker, Kraut, and Hovy}{Yang
  et~al\mbox{.}}{2017}]%
        {yang_identifying_2017}
\bibfield{author}{\bibinfo{person}{Diyi Yang}, \bibinfo{person}{Aaron
  Halfaker}, \bibinfo{person}{Robert Kraut}, {and} \bibinfo{person}{Eduard
  Hovy}.} \bibinfo{year}{2017}\natexlab{}.
\newblock \showarticletitle{Identifying {Semantic} {Edit} {Intentions} from
  {Revisions} in {Wikipedia}}.
\newblock \bibinfo{journal}{\emph{Proceedings of the 2017 Conference on
  Empirical Methods in Natural Language Processing}} (\bibinfo{year}{2017}),
  \bibinfo{pages}{2000--2010}.
\newblock


\bibitem[\protect\citeauthoryear{Zittrain, Faris, Noman, Clark, Tilton, and
  Morrison-Westphal}{Zittrain et~al\mbox{.}}{2017}]%
        {zittrain_shifting_2017}
\bibfield{author}{\bibinfo{person}{Jonathan~L. Zittrain},
  \bibinfo{person}{Robert Faris}, \bibinfo{person}{Helmi Noman},
  \bibinfo{person}{Justin Clark}, \bibinfo{person}{Casey Tilton}, {and}
  \bibinfo{person}{Ryan Morrison-Westphal}.} \bibinfo{year}{2017}\natexlab{}.
\newblock \bibinfo{title}{The {Shifting} {Landscape} of {Global} {Internet}
  {Censorship}}.
\newblock
\newblock
\urldef\tempurl%
\url{https://doi.org/10.2139/ssrn.2993485}
\showDOI{\tempurl}


\end{thebibliography}

\received{July 2022}
\received[revised]{January 2023}
\received[accepted]{March 2023}
\end{document}